\documentclass[aps,prapplied,reprint,amsmath,amssymb,superscriptaddress]{revtex4-1}
\usepackage{blindtext}
\usepackage{graphicx}% Include figure files
\usepackage{dcolumn}% Align table columns on decimal point
\usepackage{bm}% bold math
\usepackage{subfigure}
\usepackage{amssymb}
\usepackage{amsmath}
\usepackage{amsthm}
\usepackage{amsfonts}
\usepackage{mathtools}
\graphicspath{{./Pictures/}}
\usepackage[colorlinks,urlcolor=blue,linkcolor=blue,citecolor=blue]{hyperref}
\usepackage{xcolor}

\newtheorem*{lem*}{Lemma}
\newtheorem*{thm*}{Theorem}

\begin{document}
\title{Angle-robust Two-Qubit Gates in a Linear Ion Crystal}
\author{Zhubing Jia}
\email{zhubing.jia@duke.edu}
\affiliation{Duke Quantum Center, Duke University, Durham, NC 27701, USA}
\affiliation{Department of Physics, Duke University, Durham, North Carolina 27708, USA}
\author{Shilin Huang}
\affiliation{Duke Quantum Center, Duke University, Durham, NC 27701, USA}
\affiliation{Department of Electrical and Computer Engineering, Duke University, Durham, North Carolina 27708, USA}
\author{Mingyu Kang}
\affiliation{Duke Quantum Center, Duke University, Durham, NC 27701, USA}
\affiliation{Department of Physics, Duke University, Durham, North Carolina 27708, USA}
\author{Ke Sun}
\affiliation{Duke Quantum Center, Duke University, Durham, NC 27701, USA}
\affiliation{Department of Physics, Duke University, Durham, North Carolina 27708, USA}
\author{Robert F. Spivey}
\affiliation{Duke Quantum Center, Duke University, Durham, NC 27701, USA}
\affiliation{Department of Electrical and Computer Engineering, Duke University, Durham, North Carolina 27708, USA}
\author{Jungsang Kim}
\affiliation{Duke Quantum Center, Duke University, Durham, NC 27701, USA}
\affiliation{Department of Physics, Duke University, Durham, North Carolina 27708, USA}
\affiliation{Department of Electrical and Computer Engineering, Duke University, Durham, North Carolina 27708, USA}
\affiliation{IonQ, Inc., College Park, Maryland 20740, USA}
\author{Kenneth R. Brown}
\email{kenneth.r.brown@duke.edu}
\affiliation{Duke Quantum Center, Duke University, Durham, NC 27701, USA}
\affiliation{Department of Physics, Duke University, Durham, North Carolina 27708, USA}
\affiliation{Department of Electrical and Computer Engineering, Duke University, Durham, North Carolina 27708, USA}
\affiliation{Department of Chemistry, Duke University, Durham, North Carolina 27708, USA}

\begin{abstract}
In trapped-ion quantum computers, two-qubit entangling gates are generated by applying spin-dependent force which uses phonons to mediate interaction between the internal states of the ions. To maintain high-fidelity two-qubit gates under fluctuating experimental parameters, 
robust pulse-design methods are applied to remove the residual spin-motion entanglement in the presence of motional mode frequency drifts.  Here we propose an improved pulse-design method that also guarantees the robustness of the two-qubit rotation angle against uniform mode frequency drifts by combining pulses with opposite sensitivity of the angle to mode frequency drifts. We experimentally measure the performance of the designed gates and see an improvement on both gate fidelity and gate performance under uniform mode frequency offsets.
\end{abstract}

\maketitle

\section{Introduction}

Trapped atomic-ion system is a leading platform for quantum computation and simulation~\cite{Monroe2013, Brown2016}. It has several appealing features including long coherence time~\cite{Wang2017, Wang2021}, fast and efficient state preparation and measurement~\cite{Olmschenk2007, Noek2013, Harty2014}, high-fidelity single-qubit~\cite{Brown2011, Harty2014, AudeCraik2017} and two-qubit gates~\cite{Ballance2016, Gaebler2016, Wang2020, Clark2021, Baldwin2021, Srinivas2021}.
As the most challenging component, the two-qubit gates are commonly implemented by M{\o}lmer-S{\o}rensen (MS) type interactions~\cite{Molmer1999, Sorensen2000},
which utilizes the motional modes to mediate couplings between spins.
For exactly two-ion systems, two-qubit gate fidelity exceeding $99.9\%$ has been demonstrated~\cite{Ballance2016, Gaebler2016, Clark2021, Baldwin2021, Srinivas2021}. 
For longer ion chains with more complicated motional-mode spectrum,
several approaches including amplitude modulation (AM)~\cite{Roos2008, Debnath2016, Wu2018, Figgatt2019, blumel2021, Cetina2022}, frequency modulation (FM)~\cite{Leung2018, Wang2020, Kang2022, Fang2022}, phase modulation (PM)~\cite{Green2015, Lu2019, Milne2020} and multi-tone driving fields~\cite{Haddadfarshi2016, Shapira2018, Shapira2020} have been utilized to efficiently decouple spins from all motional modes, achieving $98.5\%\sim99.3\%$ fidelity with 15 ions~\cite{Egan2021} and 97.5\% fidelity with 13 ions~\cite{Wright2019}, 16 ions~\cite{Postler2022} and 25 ions~\cite{Cetina2022}.

% . 
% The MS gate 
% involves 
% phonon-mediated spin-dependent forces
% which decouple spins from motional modes and entangle the spin states at the end of the gate, performing the effective operation $\exp(i\Theta\sigma_\phi\sigma_\phi)$ with $\Theta=\pi/4$ defined as the rotation angle. 

As the circuit depth increases, 
the performance of MS gate is sensitive to slow drifts of motional mode frequencies, resulting in undesired residual spin-motion entanglement and two-qubit overrotation errors.
Previous works~\cite{Leung2018, Leung2018LongChain, Milne2020, blumel2021} have focused on minimizing both the residual spin-motion entanglement and its first-order response to unknown mode frequency drifts,
% the residual spin-motion entanglement with the presence of unknown mode frequency drifts, 
which are usually referred to as \textit{robust} gates. 
Meanwhile, few works have addressed the sensitivity of the two-qubit overrotation errors. On the circuit level, the overrotation error can be suppressed by local optimization~\cite{Debroy2018, RyanAnderson2021, Zhang2022, Majumder2022}. On the gate level, one can use single-qubit composite pulses to correct two-qubit overrotation errors~\cite{Jones2003, Merrill2014}, but due to the long two-qubit gate time it is not experimentally applicable~\cite{Zhang2022}. Refs.~\cite{Kang2021, Kang2022} applied machine learning and numerical optimization techniques to reduce the overrotation errors due to mode frequency drifts, without being able to completely remove the first-order sensitivity.
Ref.~\cite{blumel2021} proposed a scheme that eliminates such sensitivity to arbitrary order by applying different pulses on the two ions, but the scheme was not verified experimentally.

In this paper, we propose a method called \textit{A(ngle)-robust} that achieves similar goal by concatenating two different robust MS pulses.
Inheriting the robustness of negligible residual spin-motion entanglement, the two-qubit rotation angle of an A-robust gate is first-order insensitive against uniform frequency drifts on all modes. Our scheme is experimentally verified on a two-ion platform and can in principle be realized in longer ion chains. The paper is organized as follows. In Sec.~\ref{Theory}, we review the general theory of robust MS gate and discuss how its fidelity can be affected. In Sec.~\ref{ArobustTheory}, we introduce the analytic construction of A-robust pulses in two-ion chains and long ion chains and show the simulated behavior of A-robust gates comparing with that of robust gates. In Sec.~\ref{experiment} we experimentally verify that the A-robust gate outperforms the robust gate against mode frequency offsets in a two-ion chain. Finally, in Sec.~\ref{Discussions} we summarize our results and discuss further directions.

\section{Background}
\label{Theory}

\subsection{The M{\o}lmer-S{\o}rensen Gate}
The M{\o}lmer-S{\o}rensen (MS) gate entangles the spin states of two ions in an ion chain by applying a state-dependent force. 
On each target ion $j_1$ and $j_2$, 
we apply driving fields with the same Rabi frequency $\Omega(t)$ 
and opposite detunings $\pm \delta(t)$ with respect to carrier transition, where $\delta$ is close to resonance with the motional frequency.
Under Lamb-Dicke and rotating wave approximations, the effective Hamiltonian can be written as~\cite{Molmer1999, Sorensen2000}
\begin{align}
H_{\textrm MS}(t) = &\frac{\Omega(t)}{2} \sum_{j = j_1, j_2} \sum_k \eta_{k} b_j^k\left( \hat{a}_k e^{-i  \theta_k(t)} + \hat{a}_k^\dag e^{i\theta_k(t)}\right) \sigma_x^{j}
\label{Hamiltonian}
\end{align}
where $\eta_k$ is the Lamb-Dicke parameter of mode $k$, $b_j^k$ is the normalized coupling strength of ion $j$ to mode $k$, and 
\begin{align}
    \theta_k(t) =\omega_k t- \int_{0}^t \delta(t') dt'
    \label{Eq:integratedPhase}
\end{align}
is the integrated phase of the detuning between the frequency $\omega_k$ of the $k$-th mode and the driving field at time $t$. 
% \shilin{Check the sign.}
By applying Magnus expansion, at time $\tau$, the MS Hamiltonian generates a unitary evolution of the following form~\cite{Wu2018, Roos2008}
\begin{align}
    U_{\mathrm{MS}}(\tau) = \mathrm{exp}\Biggl\{&\sum_{j = j_1, j_2} \sum_k\left[\left(\alpha^k_j(\tau)\hat{a}_k^\dagger-\alpha^{k*}_{j}(\tau)\hat{a}_k\right)\sigma_x^{j}\right]\nonumber\\
    &+i\Theta(\tau)\sigma_x^{j_1}\sigma_x^{j_2}\Biggr\}
\end{align}
where 
\begin{align}
    \alpha_j^k(\tau) = \frac{\eta_{k} b_j^k}{2}\int_0^\tau{\Omega(t)e^{i\theta_k(t)}{d}t}
    \label{Eq:alpha}
\end{align}
is the displacement of mode $k$ when spin $j$ is in $(+1)$-eigenstate of $\sigma_x^j$, and
\begin{align}
    \Theta(\tau) = \frac{1}{2}\sum_k&\eta_{k}^2b_{j_1}^k
    b_{j_2}^k\int_0^\tau dt\int_0^{t} dt'\nonumber\\
    &\times\Omega(t)\Omega(t')\sin\left(\theta_k(t)-\theta_k(t')\right)
    \label{Eq:Theta}
\end{align}
is the angle of the two-qubit rotation with respect to the axis $\sigma_x^{j_1}\sigma_x^{j_2}$.

At the end of the gate, the ion spins are completely disentangled with the motional modes, i.e., $\alpha_j^k = 0$ for all ions $j$ and modes $k$. For a maximally-entangled MS gate, the rotation angle $\Theta$ should be equal to $\pi/4$. In this work we refer to maximally-entangled MS gates as $XX(\pi/4)$ gates. 
In a multi-ion chain, these conditions can be achieved by modulating either the Rabi frequency $\Omega(t)$ or the detuning $\delta(t)$ of the state-dependent driving forces in Eq.~\ref{Hamiltonian}. 
% With presence of noise, t
The gate error $\mathcal{E} = \mathcal{E}_\alpha + \mathcal{E}_{\Theta}$
% is the sum of the displacement error $\mathcal{E}_\alpha$ and rotation angle error $\mathcal{E}_{\Theta}$ which are 
can be defined as follows~\cite{Kang2022}:
\begin{align}
    \begin{split}
        \mathcal{E}_\alpha &= \sum_k
        \left(\left|\alpha_{j_1}^k\right|^2+\left|\alpha_{j_2}^k\right|^2\right),
        \\
        \mathcal{E}_\Theta &= \left(\Theta-\frac{\pi}{4}\right)^2.
        \label{gateerror}
    \end{split}
\end{align}

\subsection{Robust Pulses against Mode Frequency Drifts}
\label{Sec:RobustnessCondition}

In practice, the mode frequencies $\omega_k$ might drift to $\omega_k+\epsilon_k$ for some small $\epsilon_k$ due to miscalibration and low-frequency noise (comparing to gate speed). As a result, both the residual displacement and the rotation angle experience a first-order response
\begin{eqnarray}
    \alpha_j^k &\rightarrow& \alpha_j^k + 
    \frac{\partial \alpha_j^k}{\partial \omega_k} \epsilon_k,\\
    \Theta &\rightarrow& \Theta + 
    \sum_k\frac{\partial \Theta}{\partial \omega_k} \epsilon_k,
\end{eqnarray}
where
\begin{eqnarray}
    \frac{\partial \alpha_j^k}{\partial \omega_k} &=&
    \frac{i\eta_{k} b_j^k}{2}\int_0^\tau{\Omega(t) e^{i\theta_k(t)}t dt}
    \label{Eq:alphaDeriv}
\end{eqnarray}
and
\begin{eqnarray}
    \frac{\partial \Theta}{\partial \omega_k} &=& \frac{\eta_{k}^2 b_{j_1}^k b_{j_2}^k}{2}
    \int_0^\tau dt \int_0^t dt'\Omega(t)\Omega(t')\nonumber\\&\times& (t-t')\cos(\theta_k(t) - \theta_k(t')).
    \label{Eq:ThetaDeriv}
\end{eqnarray}
To make the MS gate behavior insensitive to mode frequency drifts, it is desirable to have pulses with vanishing first-order responses. Minimizing the residual entanglement and its first-order response to mode frequency drifts are referred to as the {robustness} condition:
\begin{itemize}
\item \textbf{Robustness:}  For all $j$ and $k$, we require that 
\begin{eqnarray}
    \alpha_j^k=0,~\frac{\partial \alpha_j^k}{\partial \omega_k} = 0.
\end{eqnarray}
\end{itemize}
Previous works have been focusing on achieving the robustness condition by applying different modulation methods, including AM~\cite{blumel2021}, FM~\cite{Leung2018, Wang2020}, PM~\cite{Green2015,Milne2020} and multi-tone driving fields~\cite{Shapira2018, Webb2018}. 
In particular, Ref.~\cite{Leung2018} shows that
% the robustness condition can be achieved 
by using a time-symmetric pulse and minimizing the absolute value of the time-averaged displacement
\begin{align}
    \overline{\alpha_j^k}=\frac{1}{\tau}\int_0^\tau \alpha_j^k(t) dt,
    \label{Eq:robustMethod}
\end{align}
the optimized pulse satisfies the robustness condition.
Meanwhile, Ref.~\cite{blumel2021} points out that using AM to eliminate the first-order sensitivity of rotation angle to drifts on each mode, i.e., $\partial\Theta/\partial\omega_k=0$ for all $k$, can only be achieved by applying different pulses on each ion. If the pulses on each ion are identical,
it is impossible to achieve $\partial\Theta/\partial\omega_k = 0$ for all $k$ by using AM~\cite{blumel2021}. Studies on minimizing $\partial\Theta/\partial\omega_k$ for all $k$ using other modulation schemes remain undeveloped.

% pulse sequences satisfying $\partial \alpha_j^k / \partial \omega_k = 0$ for all $j$ and $k$ are referred to as \textit{robust} pulses.

% by using identical pulses on two target ions, eliminating the first-order sensitivity of rotation angle to drifts on each mode, i.e., $\partial\Theta/\partial\omega_k=0$ for all $k$, is analytically proved to be impossible. While this condition can be satisfied by applying different pulses on each ion, this work indicates that making rotation angle first-order robust against drifts on each mode is not achievable in normal MS gate scheme described by Eq.~\ref{Hamiltonian}.

\section{Angle-robust Gate by Concatenating Pulses}
\label{ArobustTheory}

% To simplify the gate design,
Although minimizing $\partial\Theta/\partial\omega_k$ for each $\omega_k$ is hard to realize, we notice that in experiments, small drifts on motional modes caused by rf or dc voltage fluctuations has the form $\epsilon_k = r_k \epsilon$, indicating the mode frequency drifts are all proportional to some $\epsilon$ with certain ratio $r_k$. Based on the observation, we propose an angle-robustness condition against drifts on all modes:
% weaker condition below:

\begin{itemize}
\item \textbf{A(ngle)-robustness:}  In addition to the robustness condition, we also require that 
\begin{eqnarray}
    \sum_k\frac{\partial \Theta}{\partial \omega_k} r_k = 0.
    \label{angle}
\end{eqnarray}

\end{itemize}

Specifically, for radial mode frequency drifts in ion chains due to trap rf instability, which is a major reason of unstable radial mode frequency, uniform frequency drift  ($r_k=1$ for all $k$) is approximately a valid assumption and is used in the following discussions.

% \begin{align}
%     \frac{\partial \alpha_j^k}{\partial \epsilon_k} = 0,~
%     \sum_k\frac{\partial \Theta}{\partial \epsilon_k}\epsilon_k = 0
%     \label{angle}
% \end{align}

%However, the lack of a practical A-robust MS gate scheme can be a limiting factor in two-qubit gate performance especially when running long circuits during which overrotation error accumulates quadratically.

% In this section, we assume that the mode frequency drifts $\epsilon_k = \epsilon$ are uniform. 
% \begin{align}
%     \frac{\partial \Theta}{\partial \epsilon} = \frac{1}{2}\sum_k &\eta_{k}^2 b_{j_1}^k b_{j_2}^k \int_0^\tau dt \int_0^t dt'\nonumber\\ 
%     &\times\Omega(t)\Omega(t') (t-t')\cos(\theta_k(t) - \theta_k(t'))
% \end{align}
% is also required to construct A-robust sequences.

\subsection{Analytic Construction for Two Ions}
\label{2ionconstruct}

We first construct MS gates in a two-ion chain which involve spin-motion coupling with radial center-of-mass (COM) mode of frequency $\omega_1$ and tilt mode $\omega_2$.
The coupling strength $b_{j}^k$ of two ions on two modes are $b_1^1 = b_2^1 = b_1^2 = 1/\sqrt{2}$ and $b_2^2 = -1/\sqrt{2}$ respectively. 
By satisfying the conditions described in Eq.~(\ref{Eq:robustMethod}), we can generate a robust MS gate solution with rotation angle $\Theta={\pi}/{8}$ [$XX(\pi/8)$ gate] with some Rabi frequency $\Omega(t)$ and detuning $\delta(t)$, and we refer to this solution as $(\Omega,\delta)$. Consider another pulse with Rabi frequency $\tilde{\Omega}(t)=\Omega(t)$ and detuning \mbox{$\tilde{\delta}(t) = \omega_1 + \omega_2  - \delta(t)$}. 
Assuming $\eta_1 = \eta_2$, one can show that this new pulse $(\tilde{\Omega},\tilde{\delta})$ is also a robust $XX(\pi/8)$ gate solution whose rotation angle satisfies
\begin{eqnarray}
        % \tilde{\alpha}_j^1 = (\alpha_j^2)^\ast=0,\\
        % \tilde{\alpha}_j^2 = (\alpha_j^1)^\ast=0,\\
        % \tilde{\Theta} = \Theta=\frac{\pi}{8},\\
        % \frac{\partial \alpha_j^1}{\partial\omega_k} = \frac{\partial \tilde{\alpha}_j^2}{\partial\omega_k}=0,\\       
        \sum_{k=1,2}\frac{\partial \tilde{\Theta}}{\partial \omega_k} = - \sum_{k=1,2}\frac{\partial \Theta}{\partial \omega_k}
\end{eqnarray}
% \begin{gather}
% % \begin{split}
%             \tilde{\alpha}_j^k = (\alpha_j^k)=0,\\
%         % \tilde{\alpha}_j^2 = (\alpha_j^1)^\ast=0,\\
%         \tilde{\Theta} = \Theta=\frac{\pi}{8},\\
%         \frac{\partial \alpha_j^1}{\partial\omega_k} = \frac{\partial \tilde{\alpha}_j^2}{\partial\omega_k}=0,\\        \sum_{k=1,2}\frac{\partial \tilde{\Theta}}{\partial \omega_k} = - \sum_{k=1,2}\frac{\partial \Theta}{\partial \omega_k}.
% % \end{split}
% \end{gather}
% Therefore, the robust pulse sequence $(\tilde{\Omega}, \tilde{\delta})$ has rotation angle $\pi/8$ but with total rotation angle gradient opposite to the initial pulse $(\Omega, \delta)$. 
by using Eqs.~(\ref{Eq:alpha})(\ref{Eq:Theta})(\ref{Eq:alphaDeriv}) and (\ref{Eq:ThetaDeriv}). We then concatenate the two pulses and yield a robust $XX(\pi/4)$ gate with vanishing first-order response of rotation angle to uniform mode frequency drift, i.e., 
\begin{align}
    \Theta+\tilde{\Theta}=\frac{\pi}{4},\quad
    % ~\frac{\partial\alpha_j^k}{\partial\omega_k}=\frac{\partial\tilde{\alpha}_j^k}{\partial\omega_k}=0,
    \sum_{k=1,2}\frac{\partial(\Theta+\tilde{\Theta})}{\partial\omega_k}=0.
    \label{Robustcondition}
\end{align}
Therefore the concatenated pulse is A-robust. 
% The robustness of motion state displacement inherits from the robust $XX(\pi/8)$ solution, and the robustness of rotation angle is guaranteed by two concatenated solutions cancelling out the first-order response of total rotation angle against uniform mode frequency drift.

\begin{figure}[htbp]
     \centering
     \includegraphics[width=\linewidth]{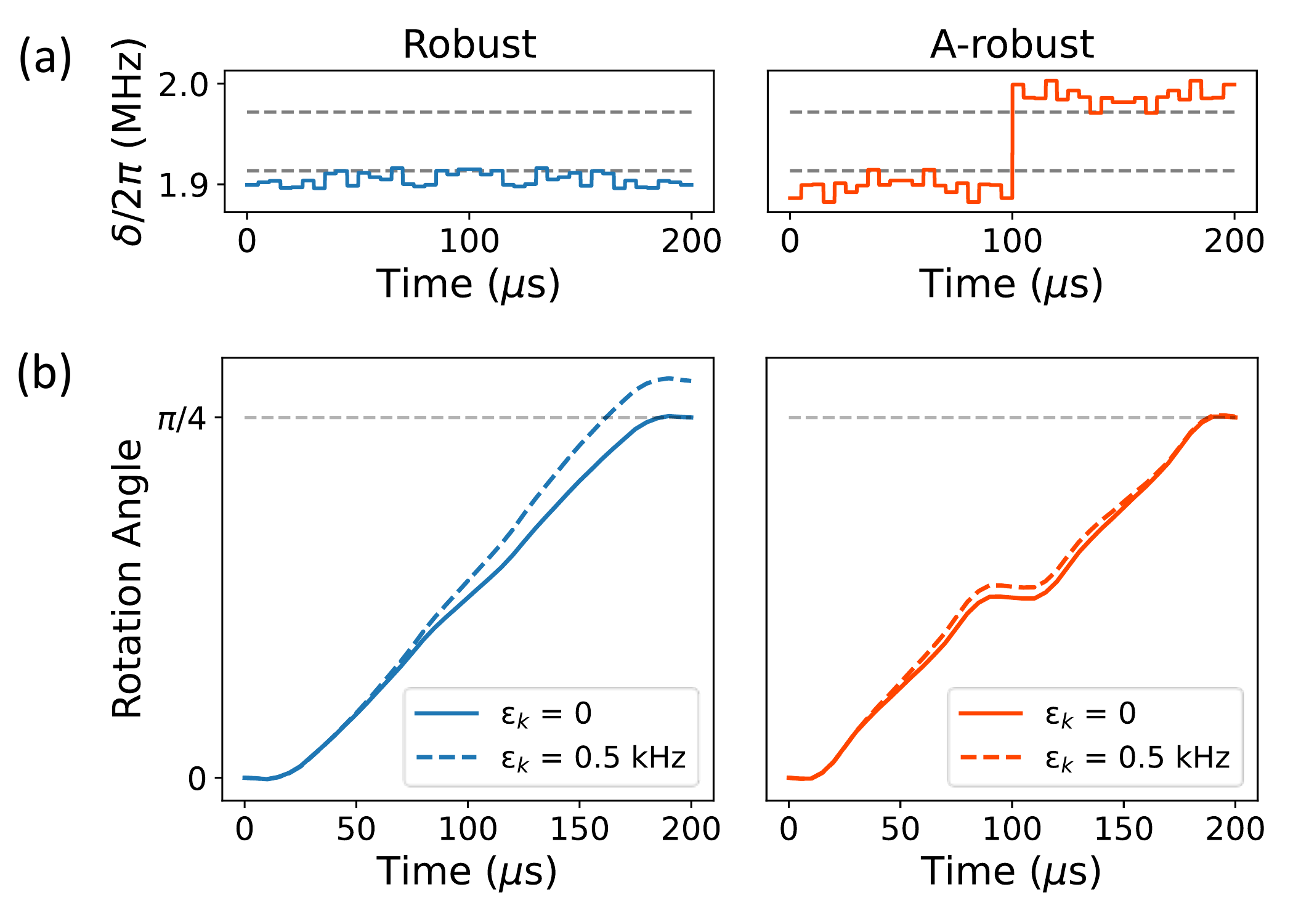}
        \caption{An example of 200-$\mu$s pulses for two-ion discrete robust (left) and A-robust FM (right) $XX\left({\pi}/{4}\right)$ gates. (a) The gray dashed lines are the radial motional-mode frequencies. The blue and orange discrete solid lines are the robust and A-robust FM pulses. The two halves of the A-robust FM pulse show symmetry with respect to the average of the two mode frequencies. Note that due to the strong coupling to the COM mode for the A-robust pulse, a low heating rate on the COM mode is beneficial. The carrier Rabi frequency for the two solutions are 68.08 kHz for A-robust and 47.21 kHz for robust FM. Under the same optimization constraints, A-robust normally requires higher power since it optimizes for a $XX\left({\pi}/{8}\right)$ gate with half of the full gate time. (b) Rotation angle of robust and A-robust pulses in (a) during the gate time. Without mode frequency drifts (solid lines), both pulses have the rotation angle reaching $\pi/4$ at the end of gate. When there exists a uniform drift of 0.5 kHz (dashed lines), the rotation angle of robust pulse has an offset from $\pi/4$, while for A-robust pulse the rotation angle is still very close to $\pi/4$.}
        \label{2ionPulses}
\end{figure}

The A-robust gate scheme can be visualized straightforwardly when concatenating two robust FM $XX(\pi/8)$ pulses that are generated using methods explained in Refs.~\cite{Wang2020,Leung2018}. 
% In particular, Ref.~\cite{Leung2018} shows that
% % the robustness condition can be achieved 
% by using a time-symmetric pulse and minimizing the absolute value of the time-averaged displacement
% \begin{align}
%     \overline{\alpha_j^k}=\int_0^\tau \alpha_j^k(t) dt
%     \label{Eq:robustMethod}
% \end{align}
% the optimized pulse can minimize $\alpha_j^k(\tau)$ and achieve the robustness condition simultaneously.
% and to simultaneously minimize $\alpha_j^k(\tau)$.
Given a robust FM $XX(\pi/8)$ solution, we flip the pulse frequencies with respect to the average of $\omega_1$ and $\omega_2$ and concatenate the two pulses. Fig.~\ref{2ionPulses} shows an example of two-ion discrete robust and A-robust FM solutions with gate time 200 $\mu$s and their rotation angle evolution with and without uniform mode frequency drifts. 
% The $XX(\pi/8)$ solution for concatenating A-robust pulse and the robust FM solution are generated using the same optimization solver~\cite{Leung2018, Wang2020}.
% Intuitively, for the rotation angle error caused by a uniform mode frequency drift, the first and second half of A-robust gate sequence experience errors with the same amplitude but opposite signs, therefore the first-order errors from the first and second half of the gate cancel each other.

\subsection{Scaling up to Long Ion Chains via Amplitude Modulation}

In long ion chains, due to the increasing number of modes and the different values of $b_j^k$, searching for two robust $XX(\pi/8)$ solutions with opposite response on rotation angle to mode frequency drift is not as simple as that in the two-ion case. Instead of searching for two robust solutions that satisfies Eq.~(\ref{Robustcondition}), we aim to find two random robust gate solutions and implement a one-step AM to adjust the Rabi frequencies of the two solutions. We use similar notations as in Sec.~\ref{2ionconstruct}: any parameter with a tilde denotes that of the second robust gate solution. Given two robust solutions $(\Omega,\delta)$, $(\tilde{\Omega},\tilde{\delta})$, their rotation angle $\Theta$, $\tilde{\Theta}$ and their gradient of rotation angle to mode frequency drift $\sum_{k}{{\partial\Theta}/{\partial\omega_k}}$, $\sum_{k}{{\partial\tilde{\Theta}}/{\partial\omega_k}}$, we can multiply each Rabi frequency by a factor $\Omega(t) \rightarrow \beta \Omega(t)$, $\tilde{\Omega}(t) \rightarrow \tilde{\beta} \tilde{\Omega}(t)$ that changes the rotation angles to 
$\Theta \rightarrow \beta^2 \Theta$, 
$\tilde{\Theta} \rightarrow \tilde{\beta}^2 \tilde{\Theta}$. The A-robust conditions (Eq.~(\ref{angle})) for a $XX(\pi/4)$ gate therefore requires $\beta$ and $\tilde{\beta}$ to satisfy the following equations
\begin{align}
\begin{split}
    \beta^2\Theta+&\tilde{\beta}^2\tilde{\Theta}=\pi/4,\\
    % \beta^2\sum_k{\frac{\partial\Theta}{\partial\omega_k}}+\tilde{\beta}^2\sum_k{\frac{\partial\tilde{\Theta}}{\partial\omega_k}}=0
    \beta^2\left(\sum_{k}{\frac{\partial\Theta}{\partial\omega_k}}\right)+&\tilde{\beta}^2\left(\sum_{k}{\frac{\partial\tilde{\Theta}}{\partial\omega_k}}\right)=0,
    \label{generalArobust}
\end{split}
\end{align}
which can be easily solved. The concatenation of two gate solutions with amplitude modulation, $(\beta\Omega, \delta)$ and $(\tilde{\beta}\tilde{\Omega}, \tilde{\delta})$, performs an A-robust gate. The two-ion A-robust pulse in Sec.~\ref{2ionconstruct} can be understood as a special case of Eq.~(\ref{generalArobust}) with $\Theta=\tilde{\Theta}=\pi/8$ and ${\sum_k{\partial\Theta}/{\partial\omega_k}}=-{\sum_k{\partial\tilde{\Theta}}/{\partial\omega_k}}$
% $\sum_k{{\partial\Theta}/{\partial\omega_k}}=-\sum_k{{\partial\tilde{\Theta}}/{\partial\omega_k}}$
, which gives the solution $\beta=\tilde{\beta}=1$. Notice that depending on the sign of $\Theta$, $\tilde{\Theta}$,
${\sum_k{\partial\Theta}/{\partial\omega_k}}$ 
% $\sum_k{{\partial\Theta}/{\partial\omega_k}}$ 
and 
${\sum_k{\partial\tilde{\Theta}}/{\partial\omega_k}}$
, Eq.~(\ref{generalArobust}) may return negative values for $\beta^2$ and $\tilde{\beta}^2$ due to different signs of $b_j^k$. This can be avoided by carefully choosing the initial-guess pulse for the pulse optimization.
% , or if individual addressing on each ion is available, by changing the phase of one individual addressing beam by $\pi$ at the beginning of the second pulse.

\begin{figure}[htbp]
     \centering
         \includegraphics[width=\linewidth]{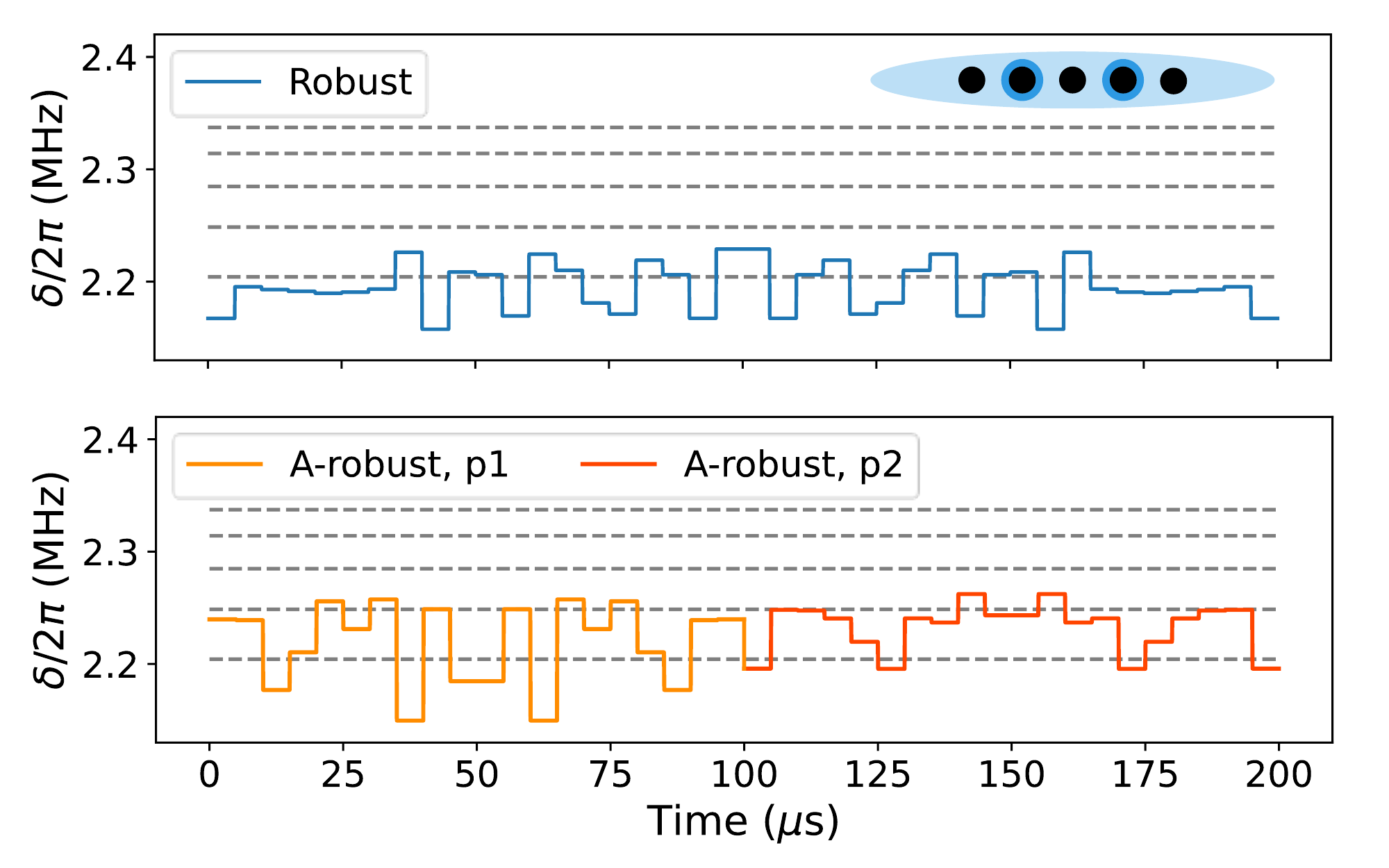}
        \caption{An example of 200-$\mu$s pulses for five-ion discrete robust and A-robust FM $XX(\pi/4)$ gates on ions 2 and 4 in a five-ion chain. The blue solid lines is the robust FM pulse, and the yellow and orange lines are the two halves of the A-robust FM pulse. After applying AM in Eq.~(\ref{generalArobust}) on the Rabi frequencies of the two halves of the A-robust pulse, the carrier Rabi frequency for the two solutions are 76.05 kHz for robust FM and 116.89 kHz/63.56 kHz for the two halves of the A-robust pulse.}
        \label{5ionPulses}
\end{figure}

Fig.~\ref{5ionPulses} shows an example of five-ion discrete robust and A-robust FM gates on ions 2 and 4 with gate time 200 $\mu$s. To find two robust pulses that gives positive solutions for Eq.~(\ref{generalArobust}), we pick the initial-guess pulses near the lowest and second-lowest mode, respectively.
% , which gives opposite signs of $b_{j_1}^k b_{j_2}^k$ from the strongest coupled mode.

\subsection{Simulated Gate Performance}
\label{Sec:2ionsim}

To compare the performances of robust and A-robust gates, we simulate the gate performance under different mode frequency drifts both for the ideal case and for the case based on our system performance~\cite{Spivey2022} using QuTiP master-equation solver~\cite{Wang2020}. 
To interpret a uniform mode frequency drift, we add detuning offsets on the pulse frequencies, since as is shown in Eq.~(\ref{Eq:integratedPhase}) they are mathematically equivalent. 
We simulate the two-qubit state populations starting from both ions at $\left|0\right>$ state after applying one MS gate with various detuning offsets (detuning scan) and the corresponding fidelity.
The simulation results shown here are of the two-ion case, and the results of the five-ion case are shown in Appendix~\ref{Sec:appxA} which has similar behavior as that of the two-ion case. For a two-ion chain, our measured heating rate is 10 quanta/s and $<1$ quanta/s on radial COM and tilt modes respectively, motional coherence time is 3 ms, and carrier coherence time is 330 ms. We emphasize that low heating rate on both modes is beneficial to A-robust gates, as it strongly couples to each mode for half of the gate time. 

\begin{figure}[htbp]
    \centering
     \includegraphics[width=\linewidth]{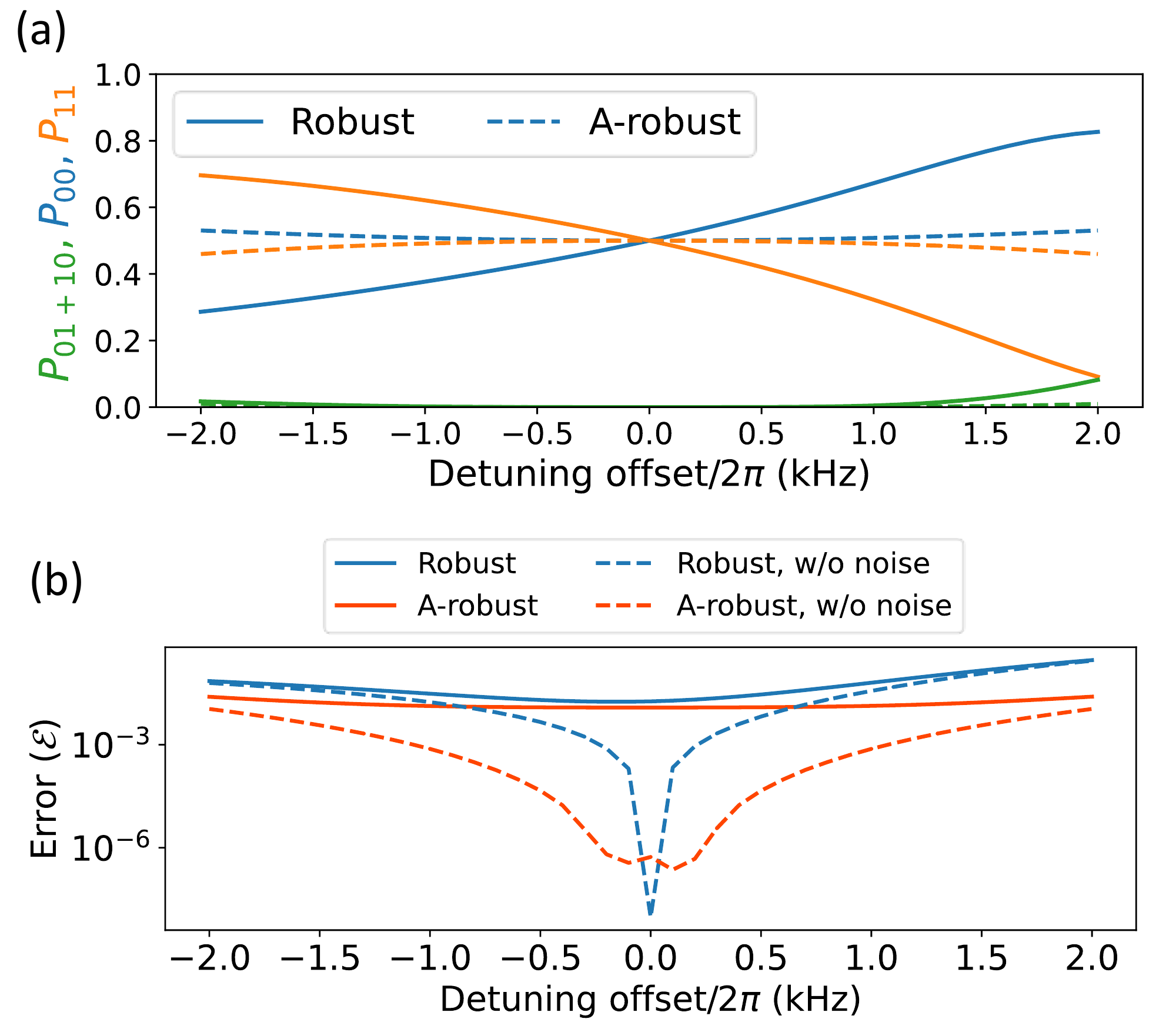}
        \caption{Simulated detuning scan without stochastic noise (a) and fidelity with/without stochastic noise (b) for the robust and A-robust pulses in Fig.~\ref{2ionPulses}. (a) The A-robust gate inherits the residual displacement robustness, exhibiting flat odd-parity populations ($P_{01+10}$) within around 2 kHz detuning range. The even-parity  populations ($P_{00}$ and $P_{11}$) for A-robust has no first-order response on detuning, thus are both closer to the ideal value 0.5 in the presence of detuning offset. 
        (b) Without stochastic noise, in the presence of detuning offsets, the error of the A-robust gate is orders of magnitude smaller than that of the robust gate. Note that the small gate errors at zero detuning offset are caused by imperfection of the numerically-optimized solutions. With stochastic noise, the A-robust gate still outperforms the robust gate. 
        In a system with less stochastic noise where gate error less than $10^{-2}$ can be reached, the advantage of the A-robust gate becomes more significant (see Appendix~\ref{Sec:appxC}).}
    \label{fidelity}
\end{figure}

Assuming perfect Rabi frequency calibration, Fig.~\ref{fidelity} shows the two-qubit population of detuning scan and fidelity of the robust and A-robust FM gate pulses in Fig.~\ref{2ionPulses} with various detuning offsets. Here the fidelity is defined as the Bell state fidelity, including errors from both residual displacement and rotation angle. The A-robust gates inherits the residual displacement robustness from robust gates, while having the first-order response of total rotation angle to detuning offset removed. Due to the robustness of rotation angle, within 2 kHz static detuning offset, the A-robust gate shows better tolerance to detuning offsets than the robust gate. 

\begin{figure}[htbp]
    \centering
    \includegraphics[width=\linewidth]{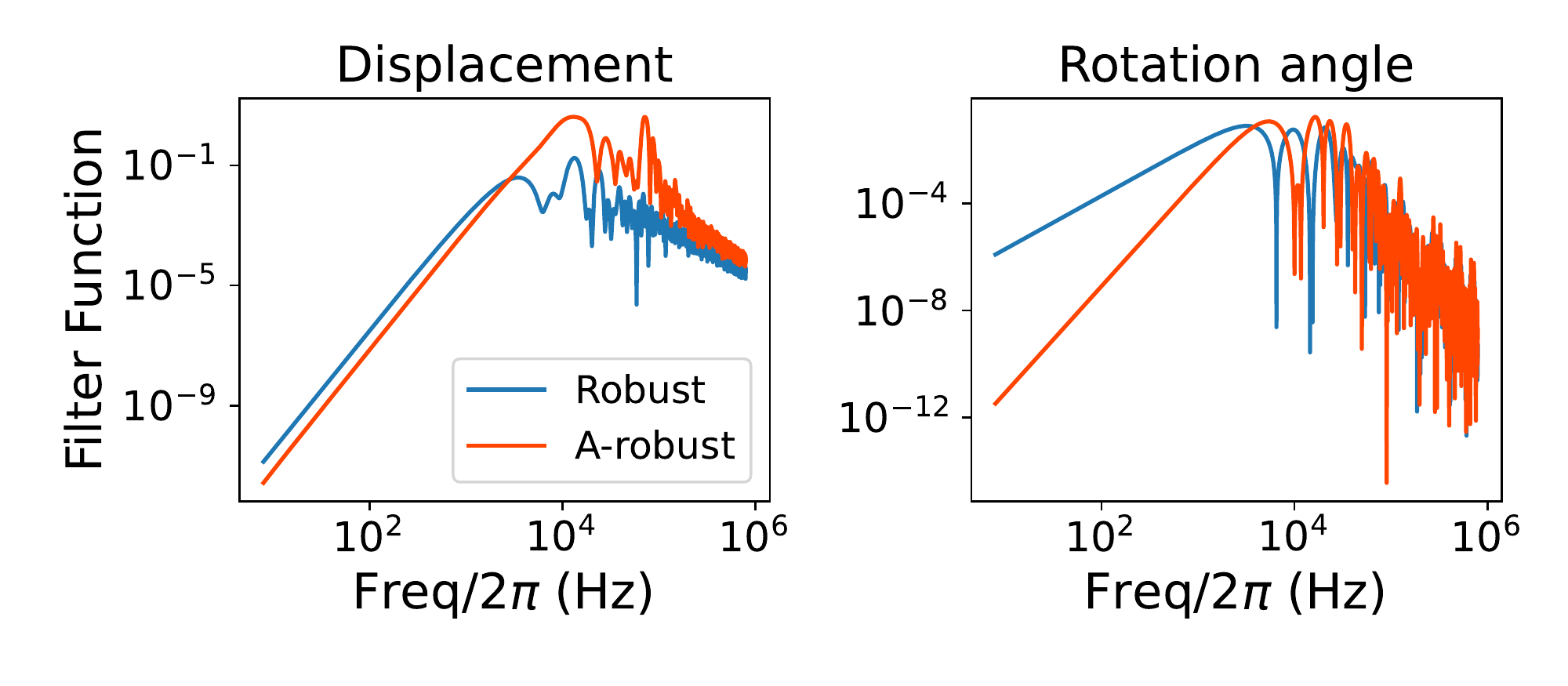}
    \caption{The FFs of the pulses in Fig.~\ref{2ionPulses} for residual displacement and rotation angle. Robust and A-robust pulses have similar FFs for residual displacement, and A-robust has lower response of rotation angle to noise with frequency below 10$^3$ Hz.}
    \label{filterfunc}
\end{figure}

We further investigate the filter function (FF) of A-robust pulses for both residual displacement and rotation angle. Details about the FF formalism and how it applies to gate robustness are discussed in Refs.~\cite{Green2015, Milne2020, Ball2021, Kang2022}. Assuming the mode frequencies experience a uniform time-varying fluctuation $\omega_k \rightarrow \omega_k+\delta(t)$, the gate errors in Eq.~\ref{gateerror} are given by

\begin{align}
    \mathcal{E}_\nu = \int_{-\infty}^{\infty}df\frac{S_\delta(f)}{f^2}F_\nu(f),~ \nu = \alpha,\Theta
\end{align}
where $S_\delta(f)$ is the power spectral density of $\delta(t)$ and
\begin{align}
    F_\alpha(f) =& \sum_k\left[\left(b^k_{j_1}\right)^2+\left(b^k_{j_2}\right)^2\right]\left|\frac{\eta_k^2}{2}\int_0^{\tau}dt\Omega(t)e^{i(2\pi ft-\theta_k(t))} \right|^2
\\
    F_{\Theta}(f) = & \left|\int_0^\tau dt \int_0^{t} dt' \left(e^{2\pi ift}-e^{2\pi ift'}\right)\Omega(t)\Omega(t')\right.\nonumber\\
    &\times\left.\sum_k\frac{\eta_k^2}{2}b^k_{j_1}b^k_{j_2}\cos\left[\theta_k(t)-\theta_k(t')\right]\right|^2
\end{align}
are the FFs for residual displacement and rotation angle~\cite{Kang2022}. Fig.~\ref{filterfunc} compares the displacement and rotation angle FFs of the robust and A-robust pulses in Fig.~\ref{2ionPulses}. The two displacement FFs have similar behaviors, indicating the A-robust gate maintains the robustness against displacement error as in the robust-FM gate. The rotation angle FFs demonstrate the improved rotation angle robustness against low-frequency noise of the A-robust gate. 
Below $10^4$ Hz, rotation angle FF of the A-robust gate shows a steeper slope, indicating removal of first-order response of the rotation angle to static mode frequency offsets~\cite{Kang2022}. 

\section{Experiments}
\label{experiment}

We experimentally measure the gate performance of robust and A-robust gates for the two-ion case. The experiment is conducted in a compact cryogenic trapped-ion system~\cite{Spivey2022}. A two-ion $^{171}$Yb$^{+}$ chain with a separation of about 5 $\mu$m is confined in a micro-fabricated linear radio-frequency Paul trap~\cite{Revelle2020} with radial motional modes $\omega\sim(2\pi)2.0$ MHz. 
The qubit levels are encoded in the hyperfine ground states $\left|\downarrow\right>\equiv~^2$S$_{1/2}\left|F=0, m_F=0\right>$ and $\left|\uparrow\right>\equiv~^2$S$_{1/2}\left|F=1, m_F=0\right>$ with a frequency splitting of $(2\pi)12.642821$ GHz~\cite{Olmschenk2007}. The qubit rotations and ion-motion coupling transitions are driven by stimulated Raman transitions using two counter-propagating mode-locked 355 nm picosecond-pulsed laser beams. One global beam is shined on both ions and two individual addressing beams are tightly focused onto each ion. The amplitude, frequency and phase of Raman beams are controlled by a Xilinx Zync UltraScale+ Radio Frequency System-on-Chip (RFSoC) board using firmware designed by Sandia National Laboratories~\cite{RFSoC}.

The robust FM pulses are determined by a numerical optimizer which minimizes the residual displacement while guaranteeing its robustness to detuning offsets. Given the gate time, motional-mode frequencies, and constraints on the target Rabi frequency as input parameters, we let the optimizer search for a $XX(\pi/8)$ solution with half of the gate time. Then we either repeat the solution pulse twice to generate a robust $XX(\pi/4)$ solution or generate an A-robust $XX(\pi/4)$ solution using the construction method described in Sec. \ref{2ionconstruct}, such that the improvement of A-robust is not affected by other parameters such as different Rabi frequencies. The Bell-state fidelity is extracted by measuring the odd-parity population ($P_{01+10}$) and parity contrast ($1-2|\rho_{00,11}|$)~\cite{Leibfried2003}, which reflect the displacement error and overrotation error respectively. In Appendix~\ref{Sec:appxB} we show the measured Bell-state fidelity of a FM $XX(\pi/4)$ solution, which is lower than that of either repeated $XX(\pi/8)$ robust gate or A-robust gate.

\begin{figure}[htbp]
    \centering
    \includegraphics[width=\linewidth]{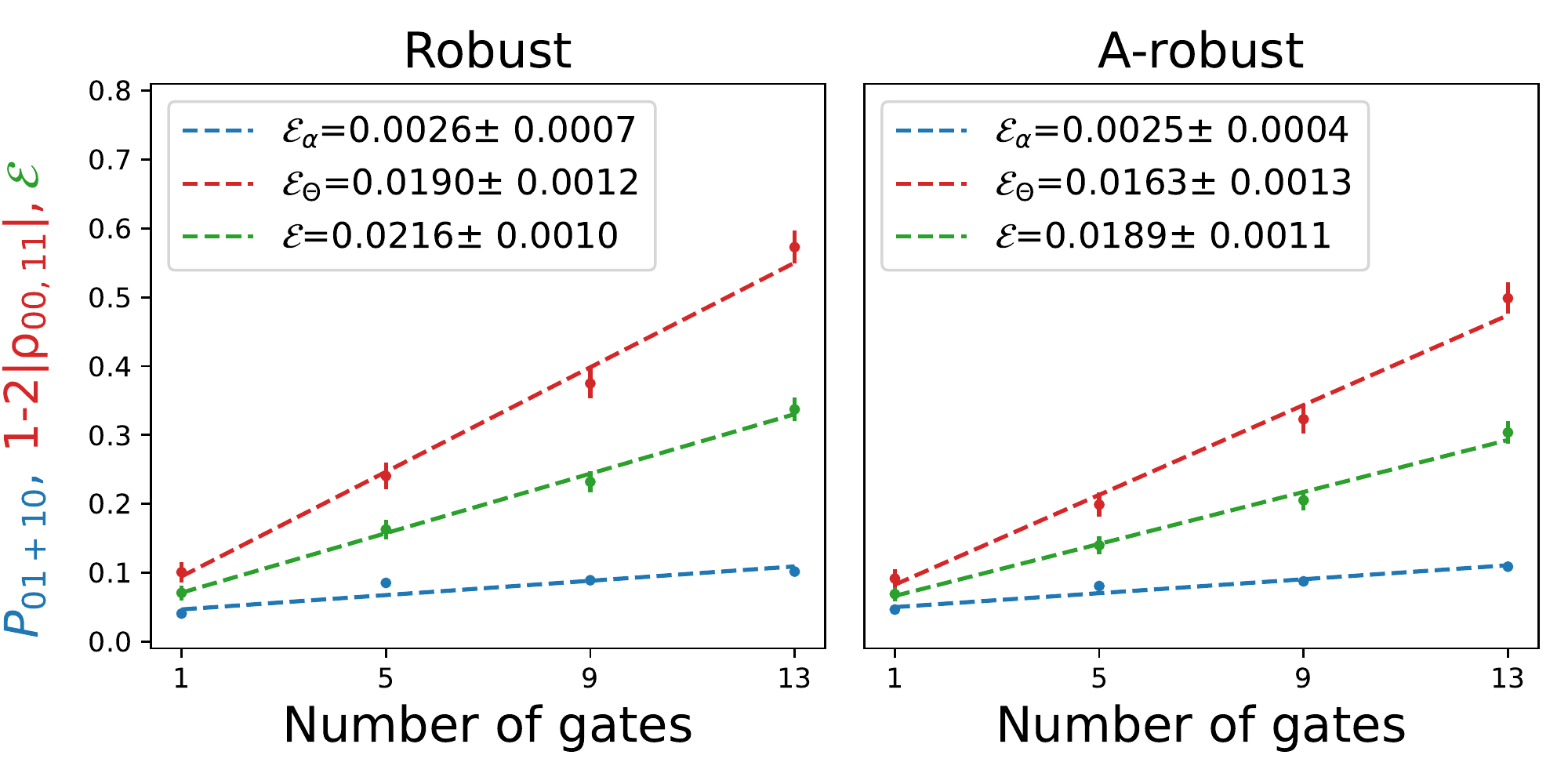}
    \caption{Gate-error measurement of the robust and A-robust gate by applying repeated gate sequences. The blue, red and green dots and lines are the odd-parity population of $\left|01\right>$ and $\left|10\right>$ states, the parity contrast $1-2|\rho_{00,11}|$, and the measured gate error $\mathcal{E}$. Both gates have similar odd-parity populations, and the parity contrast of the A-robust gate outperforms that of the robust gate.}
    \label{2ionFid_data}
\end{figure}

We compare the performance of the robust and A-robust FM MS-gate pulses with fixed gate time 270 $\mu$s, by measuring the MS-gate fidelity and checking the gate performance under detuning offsets. The Bell-state fidelity is measured using the method in Ref.~\cite{Wang2020}: We initialize the qubit to $\left|00\right>$ state, then apply a sequence of 1, 5, 9 and 13 robust or A-robust FM gates and measure the final state fidelity. Finally we extract the gate fidelity from a linear fit of final state fidelity versus gate number. The measured MS-gate fidelity is $97.84(1)\%$ for the robust gate and $98.11(1)\%$ for the A-robust gate, as is shown in Fig.~\ref{2ionFid_data}. Notice that both gates have similar $\mathcal{E}_\alpha\approx 0.25\%$, but the A-robust gate has better parity contrast, which leads to $0.27\%$ overall fidelity improvement, indicating a more robust rotation angle under system noise. We expect a more significant improvement in systems where the mode frequency drift is dominated by slow coherent noise instead of stochastic noise.

\begin{figure}[htbp]
    \centering
    \includegraphics[width=\linewidth]{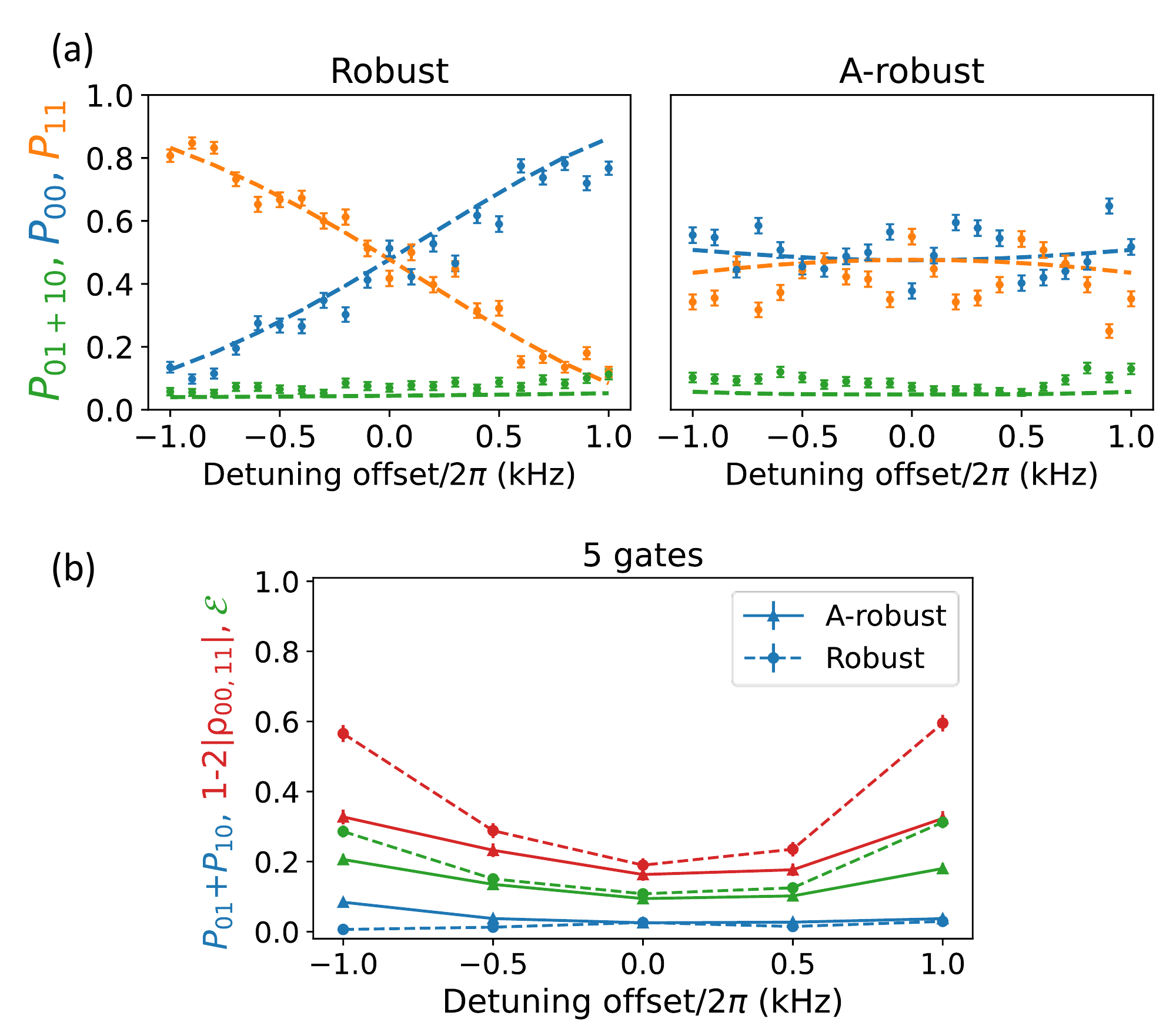}
    \caption{(a) Experimental data (points) and simulation (dashed lines) of $P_{01+10}$, $P_{00}$ and $P_{11}$ after applying five concatenated robust or A-robust gates, with detuning offsets ranging from -1 kHz to 1 kHz. The $P_{00}$ and $P_{11}$ curves of the robust gate has a linear dependency on detuning offset, while for the A-robust gate the first-order response is removed. (b) Measured $P_{01+10}$, $1-2|\rho_{00,11}|$ and error $\mathcal{E}$ of five concatenated robust and A-robust gates with detuning offsets. Overall the A-robust gate has better fidelity in the presence of detuning offsets.}
    \label{2ionDetuning_data}
\end{figure}

We further investigate the gate performance under mode frequency drifts by introducing detuning offsets to the pulses and measuring the fidelity of the robust or A-robust gate. Fig.~\ref{2ionDetuning_data} shows the detuning scan results and Bell state fidelity of five concatenated robust and A-robust gates under detuning offsets from $-1$ to 1 kHz. Both robust and A-robust gates have $P_{01+10}$ close to zero and robust against detuning offsets. Also, the A-robust gate has its even-parity populations more robust against detuning offsets, which implies a robust rotation angle.

\section{Discussions and Outlook}
\label{Discussions}

The idea of A-robust strategy can be incorporated with various other optimization schemes and the optimization process can be adjusted based on realistic situations. A few possible future directions are listed here.

\subsection{Combining A-robust with other Optimization Schemes}

An A-robust gate inherits robustness of the residual displacement from robust gates, but what it can inherit is not limited to this: any pulse optimized for a certain feature can be concatenated to form an A-robust pulse with the same feature. 

\begin{figure}[htbp]
    \centering
    \includegraphics[width=\linewidth]{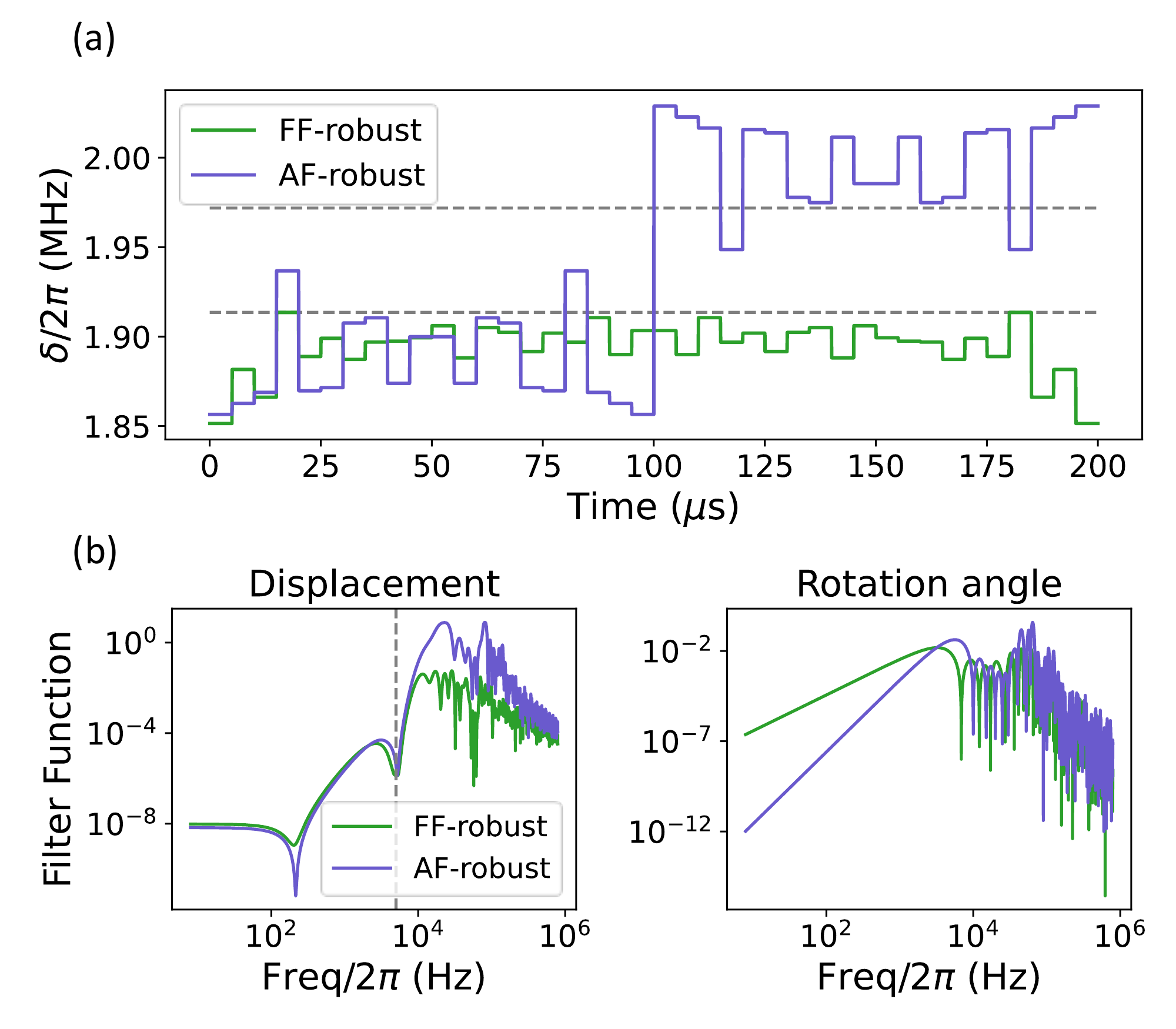}
    \caption{(a) An example of 200-$\mu$s pulses for two-ion discrete FF-robust and AF-robust FM $XX(\pi/4)$ pulses. The carrier Rabi frequencies for the FF- and AF-robust pulses are 73.65 kHz and 100.36 kHz respectively. (b) The FFs of the pulses in (a). The pulses are optimized such that the displacement FF is suppressed near 5 kHz (dashed line).}
    \label{2ionAFrobust}
\end{figure}

Here we take filter-function robust (FF-robust) optimization protocol as an example~\cite{Kang2022}. Apart from suppressing low-frequency noise, the FF-robust optimization scheme searches for pulse that is resilient to noise with certain frequency. By concatenating two FF-robust pulses we can generate an A-robust gate (``AF-robust''), which maintains both residual-displacement and rotation angle robustness in low frequency domain and certain high frequency range as designed. Fig.~\ref{2ionAFrobust} shows an example of two-ion FF-robust and AF-robust pulses and their filter functions that are designed to suppress noise of 5 kHz. In Appendix~\ref{Sec:appxC} we show the fidelity of the robust, A-robust, FF-robust and AF-robust gates with different detuning offsets and noise parameters. We see that the AF-robust gate gives overall the best performance against detuning offset and stochastic noise.

\subsection{Non-Uniform Drifts on Motional Modes}

If axial modes instead of radial modes are used for applying gates, or the mode frequency instability is dominated by fluctuating dc voltages, the uniform-drift assumption fails. Therefore in general we aim to satisfy the angle-robustness condition in Eq.~\ref{angle} with different but fixed values of $r_k$ for each mode $k$.
% Generally, if the mode frequency drift is caused by noise on the trap's RF or DC voltages, each mode $\omega_k$ experiences a drift $\epsilon_k=r_k\epsilon$, indicating the response of each mode differs up to a constant. Therefore the rotation angle robustness condition in Eq.~\ref{angle} can be rewritten as
% \begin{align}
%     \sum_k\frac{\partial \Theta}{\partial \omega_k}r_k = 0
% \end{align}
% which multiplies the gradient of rotation angle to mode frequency drift by $r_k$ for each $\omega_k$. 
This can be taken into account when solving the factors $\beta$ and $\tilde{\beta}$ in Eq.~\ref{generalArobust} by changing it into the form
\begin{align}
\begin{split}
    \beta^2\Theta&+\tilde{\beta}^2\tilde{\Theta}=\pi/4,\\
    % \beta^2\sum_k{\frac{\partial\Theta}{\partial\omega_k}}+\tilde{\beta}^2\sum_k{\frac{\partial\tilde{\Theta}}{\partial\omega_k}}=0
    \beta^2\left(\sum_{k}{\frac{\partial\Theta}{\partial\omega_k}r_k}\right)&+\tilde{\beta}^2\left(\sum_{k}{\frac{\partial\tilde{\Theta}}{\partial\omega_k}r_k}\right)=0.
    % \label{generalArobust}
\end{split}
\end{align}
Thus the A-robust optimization scheme holds for non-uniform drift, providing that the A-robustness condition holds (comparatively small noise) and that the proportionalities of mode frequency drifts is known.

\subsection{Arbitrary-order Suppression of Rotation Angle Error to Detuning Offsets}

The A-robust concatenation protocol can be extended to suppressing the response of rotation angle error to detuning offsets up to arbitrary order. To realize a $n$-th order suppression, we require $n+1$ different robust solutions $(\Omega_i,\delta_i)$ for $i=1,\cdots,n+1$, each with total rotation angle $\Theta_i$ and gradient over mode frequency drift up to $n$-th order, i.e., $\sum_k{\partial^j \Theta_i}/{(\partial \omega_k)^j}$ where $j=1,\cdots,n$. Then we apply AM to each robust solution by multiplying each Rabi frequency with a factor $\beta_i$ which is determined by the following set of linear equations:

\begin{align}
\begin{split}
    \sum_{i=1}^{n+1}{\beta_i^2\Theta_i} &= \frac{\pi}{4},\\
    \sum_{i=1}^{n+1}{\beta_i^2\left(\sum_k\frac{\partial^j\Theta_i}{(\partial\omega_k)^j}\right)} &= 0, j=1,\cdots,n.
\end{split}
\end{align}

The concatenation of $(\beta_i\Omega_i, \delta_i),~i=1,\cdots,n+1$ gives the total rotation angle equal to $\pi/4$ and the gradient over mode frequency drift up to $n$-th order equal to zero, leading to a $n$-th order A-robust $XX(\pi/4)$ gate.

\section{Conclusion}

In this paper, we show that by concatenating two robust MS-gate solutions and applying AM to their Rabi frequencies, we can demonstrate a gate scheme which has both residual displacement and rotation angle robust against static offsets or slow drifts in the mode frequencies. Experimentally we verify the rotation angle robustness of the A-robust gate against detuning offsets in a compact cryogenic trapped-ion system and achieve two-qubit gate fidelity of $98.11(1)\%$. We note that the A-robust strategy described in this work is applicable to any two-qubit gate scheme that involves geometric phase, such as MS gates, light-shift gates and laser-free gates~\cite{Solano1999, Milburn2000,Leibfried2003,Baldwin2021,Clark2021,Srinivas2021}.

When the gate time is fixed, the required Rabi frequency (or optical power) for A-robust gates is normally higher than that of robust gates. With a well-designed pulse optimizer, the required Rabi frequency can be constrained at the cost of solution imperfection.

One limitation of the A-robust strategy is that it cannot compensate the rotation angle error due to optical power miscalibration or fluctuation, which remains challenging using only MS-type interactions~\cite{Kang2022}. A recent work~\cite{Shapira2022} provides a possible solution to power drifts by exciting second-order spin-motion coupling, thus implementing spin-dependent squeezing, at the cost of larger Rabi frequencies.

\begin{acknowledgements}

We thank Pak Hong Leung and Ye Wang for useful discussions.
This work was supported by the Office of the Director of National Intelligence, Intelligence Advanced Research Projects Activity through ARO Contract W911NF-16-1-0082, the National Science Foundation Expeditions in Computing Award 1730104, the National Science Foundation STAQ Project Phy-181891, and the U.S. Department of Energy, Office of Advanced Scientific Computing Research QSCOUT program, Office of Advanced Scientific Computing Research Award No. DE-SC0019294 (trapped-ion tomography), DOE Basic Energy Sciences Award No. DE-0019449 (spin-motion pulses) and ARO MURI Grant No. W911NF-18-1-0218 (distributed measurement protocol).

% The QSCOUT and tomography - Are they the same one?

\end{acknowledgements}

~

\textit{Note added.} --- After completion of this work, we became aware of a preprint appeared on arXiv recently~\cite{Ruzic2022} which also aims to eliminate the first-order response of two-qubit rotation angle to uniform frequency drift by using AM and targeting the lowest two motional modes.

\appendix

\section{Five-ion Simulation Results}
\label{Sec:appxA}

In Sec.~\ref{Sec:2ionsim} we show the simulation results of the robust and A-robust gates for the two-ion case. For multi-ion cases, the simulation results are similar to those of the two-ion case, and here we show the results of the five-ion pulses in Fig.~\ref{5ionPulses}.

\begin{figure}[htbp]
     \centering
    \includegraphics[width=\linewidth]{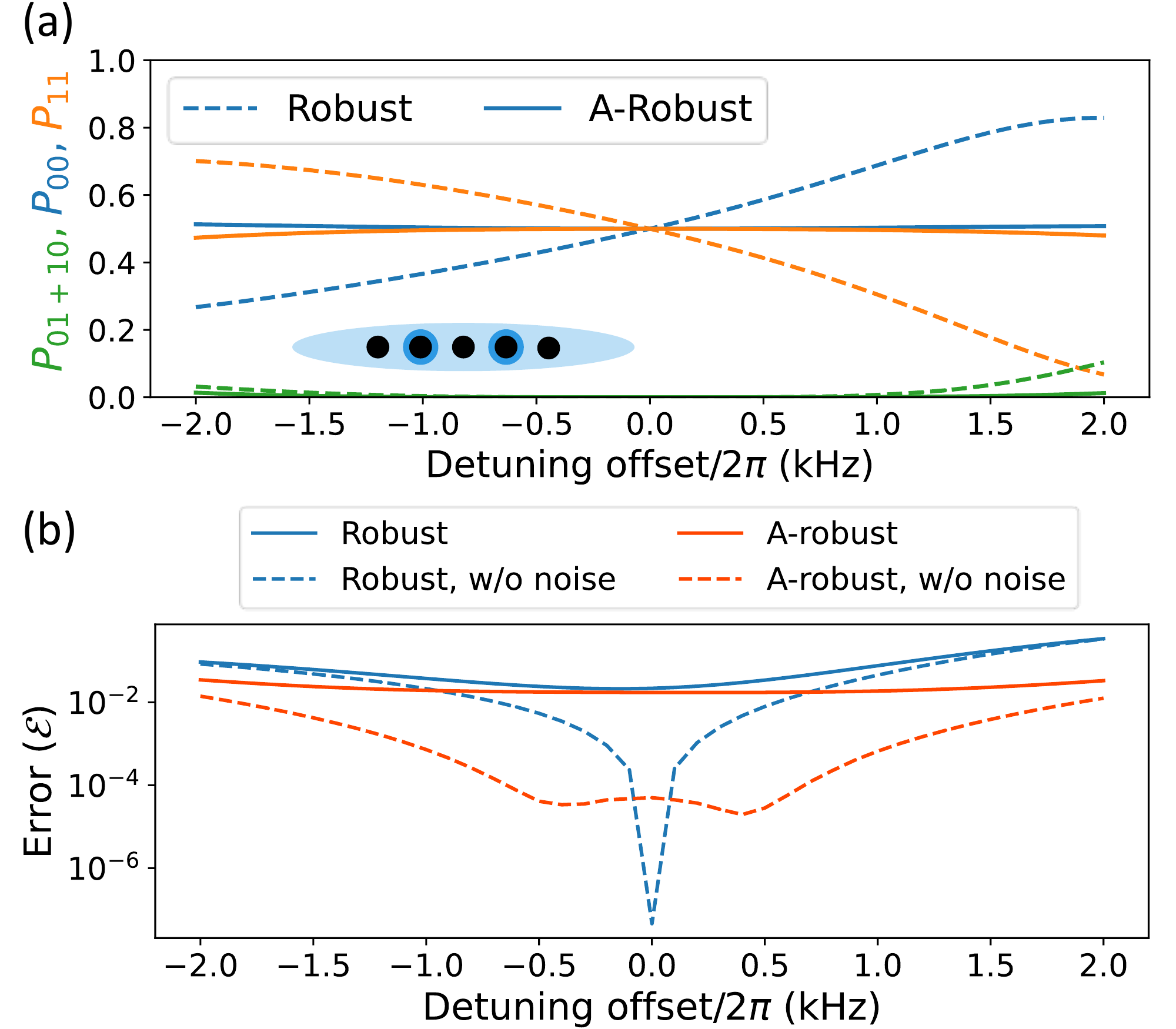}
        \caption{Simulated results of the robust and A-robust pulses in Fig.~\ref{5ionPulses}, which are for a five-ion chain. We simulate (a) detuning scan and (b) gate error with different detuning offsets. Here we use 3 ms motional coherence time and 330 ms carrier coherence time. The motional heating rate used here are 50 quanta/s on the COM mode, 5 quanta/s on the second highest mode, and 0 quanta/s on all other modes.}
        \label{5ionFidelity}
\end{figure}

Fig.~\ref{5ionFidelity} shows the simulated detuning scan and gate errors of the pulses on ions 2 and 4 of a five-ion chain, shown in Fig.~\ref{5ionPulses}, assuming perfect Rabi-frequency calibration. Similar to two-ion simulation results, the A-robust pulse has flat $P_{00}$ and $P_{11}$ with respect to detuning offsets, whereas the robust pulse holds a first-order dependence of $P_{00}$ and $P_{11}$ to detuning offsets. When looking at the fidelity, at zero detuning the robust gate has higher fidelity than the A-robust gate, mainly because each half segment of the A-robust pulse uses half the number of discrete steps compared to that of the robust pulse. When there exists small detuning offset or stochastic noise, the five-ion A-robust gate outperforms the robust gate.

\begin{figure}[htbp]
     \centering        \includegraphics[width=\linewidth]{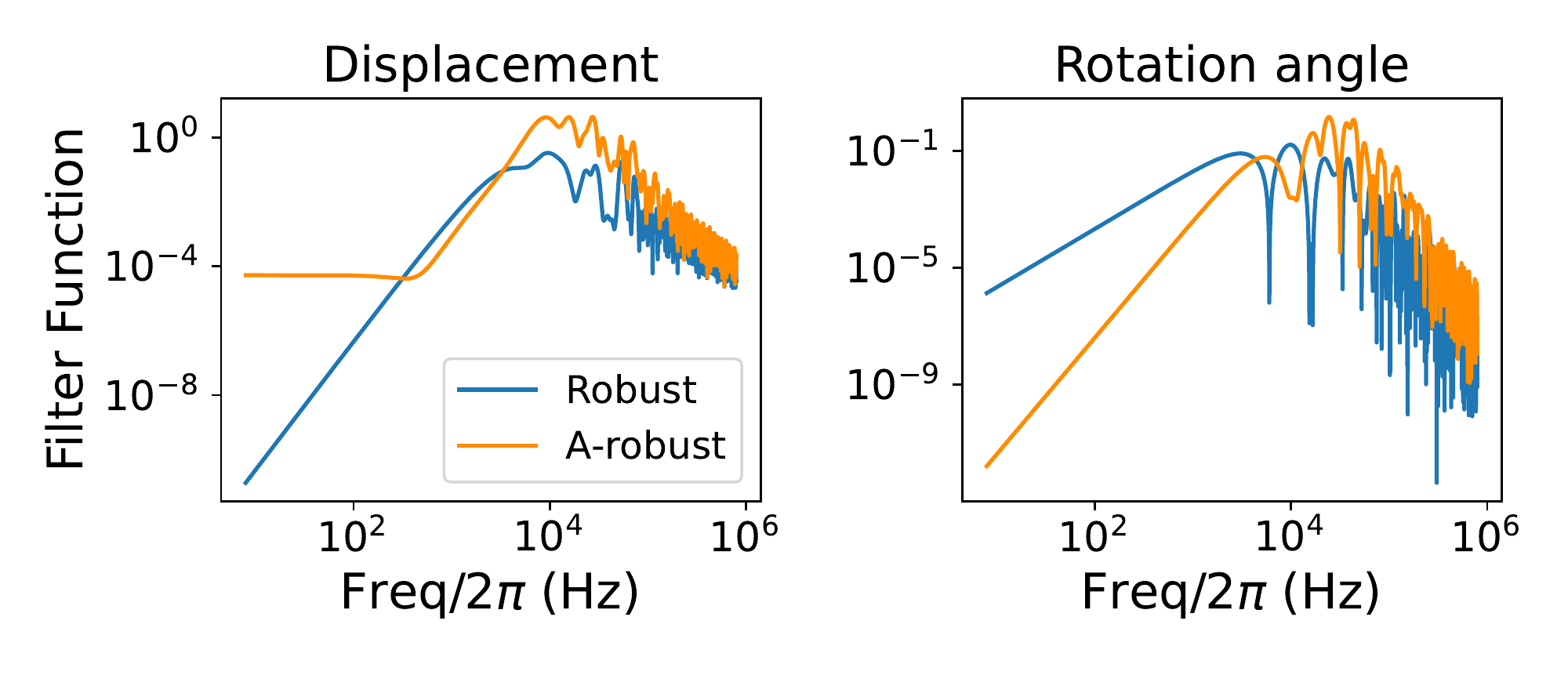}
        \caption{The FFs for the five-ion robust and A-robust pulses in Fig.~\ref{5ionPulses}. The displacement FF of the A-robust pulse converges to a small but nonzero value at low frequencies, due to imperfection of the pulse solution. Similar to the two-ion case, the rotation angle FF has a steeper slope when the frequency is below 10$^{4}$ Hz.}
        \label{5ionFF}
\end{figure}

We further check the filter functions of the five-ion robust and A-robust pulses. Fig.~\ref{5ionFF} shows the FFs of the pulses in Fig.~\ref{5ionPulses}. The rotation angle FF shows a similar behavior as in Figs.~\ref{filterfunc} and \ref{2ionAFrobust}: steeper slope at frequencies below 10$^4$ Hz, indicating the first-order suppression of overrotation error.

\section{Robust $XX(\pi/4)$ Gate Behavior}
\label{Sec:appxB}

\begin{figure}[htbp]
    \centering
    \includegraphics[width=0.8\linewidth]{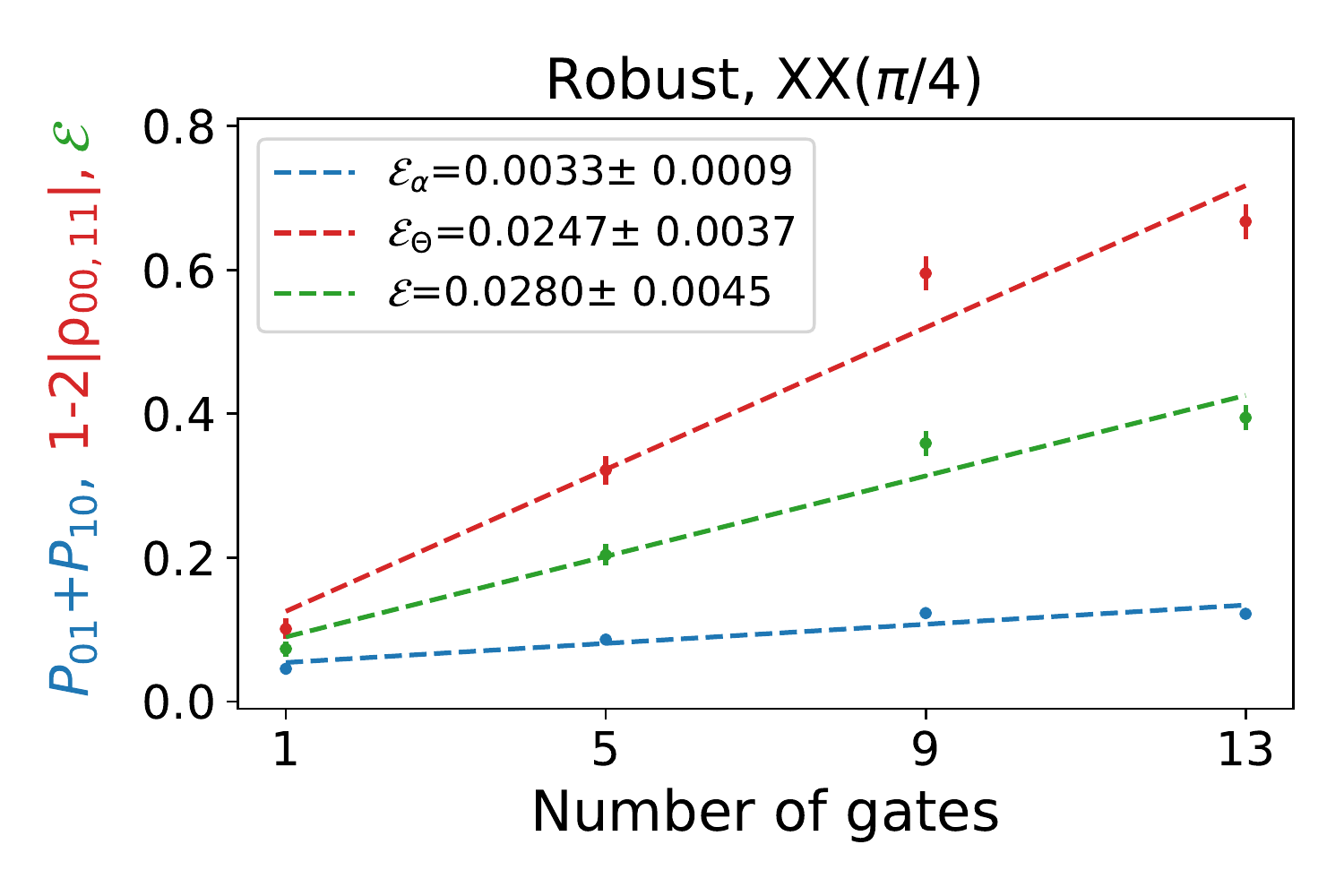}
    \caption{Gate-error measurement of the robust $XX(\pi/4)$ pulse. Here the pulse is generated by directly searching for a robust $XX(\pi/4)$ solution, instead of repeating two robust $XX(\pi/8)$ pulses.}
    \label{2ionFid_data_direct}
\end{figure}

As a comparison with A-robust gate and repeated robust $XX(\pi/8)$ gate, we directly optimize for a robust $XX(\pi/4)$ solution with the same gate time 270 $\mu$s then measure the gate fidelity using the same approach as in Sec.~\ref{experiment}. Fig.~\ref{2ionFid_data_direct} shows the measurement result, giving a gate fidelity of 97.20(4)\%. The measured robust $XX(\pi/4)$ fidelity is lower than either the A-robust gate or the repeated robust $XX(\pi/8)$ gate.

\section{Comparing Robust, A-robust, FF-robust and AF-robust Performances}
\label{Sec:appxC}

We numerically verify that when comparing robust pulse, A-robust pulse, FF-robust pulse, and concatenation of two FF-robust pulses using the A-robust strategy (``AF-robust''), the AF-robust pulse behaves overall the best against detuning offsets. 

\begin{figure*}[htbp]
     \centering         
     \includegraphics[width=\linewidth]{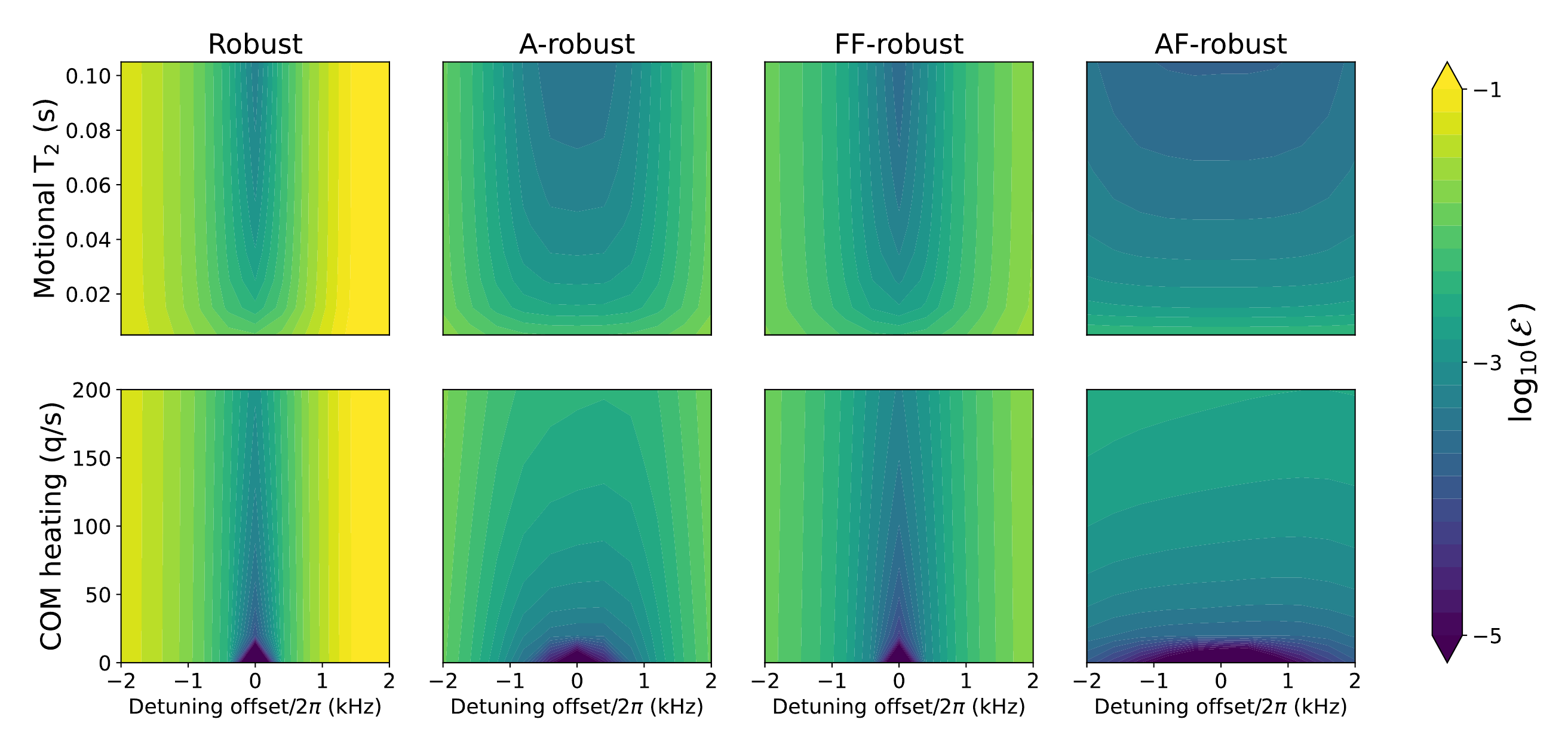}
    \caption{The simulated gate errors of one robust and A-robust pulses in Fig.~\ref{2ionPulses} and FF-robust and AF-robust pulses in Fig.~\ref{2ionAFrobust}. The first row simulates the gate error with the motional coherence time $T_2$ varying from 5 ms to 100 ms. The second row simulates the gate error with the COM-mode heating rate varying from 0 q/s to 200 q/s and the tilt mode heating rate set as 1/10 of the COM-mode heating rate. The A-robust and AF-robust pulses have larger tolerance against detuning offsets, and the AF-robust pulse overall gives the best gate fidelity in the presence of detuning offsets and stochastic noise.}
    \label{2ionFidelityComp}
\end{figure*}

Fig.~\ref{2ionFidelityComp} shows the simulated gate errors of all four pulses in Figs.~\ref{2ionPulses} and \ref{2ionAFrobust} with various detuning offsets. Comparing with the robust and FF-robust pulses, the A-robust and AF-robust pulses show larger tolerance against detuning offsets. In particular, the AF-robust pulse further suppresses the error to $<10^{-4}$ without noise and $<10^{-2}$ with heating and motional dephasing noise, at the cost of requiring the highest Rabi frequency.

\bibliographystyle{apsrev4-1} % Tell bibtex which bibliography style to use
\bibliography{./mybib}

\begin{thebibliography}{53}%
\makeatletter
\providecommand \@ifxundefined [1]{%
 \@ifx{#1\undefined}
}%
\providecommand \@ifnum [1]{%
 \ifnum #1\expandafter \@firstoftwo
 \else \expandafter \@secondoftwo
 \fi
}%
\providecommand \@ifx [1]{%
 \ifx #1\expandafter \@firstoftwo
 \else \expandafter \@secondoftwo
 \fi
}%
\providecommand \natexlab [1]{#1}%
\providecommand \enquote  [1]{``#1''}%
\providecommand \bibnamefont  [1]{#1}%
\providecommand \bibfnamefont [1]{#1}%
\providecommand \citenamefont [1]{#1}%
\providecommand \href@noop [0]{\@secondoftwo}%
\providecommand \href [0]{\begingroup \@sanitize@url \@href}%
\providecommand \@href[1]{\@@startlink{#1}\@@href}%
\providecommand \@@href[1]{\endgroup#1\@@endlink}%
\providecommand \@sanitize@url [0]{\catcode `\\12\catcode `\$12\catcode
  `\&12\catcode `\#12\catcode `\^12\catcode `\_12\catcode `\%12\relax}%
\providecommand \@@startlink[1]{}%
\providecommand \@@endlink[0]{}%
\providecommand \url  [0]{\begingroup\@sanitize@url \@url }%
\providecommand \@url [1]{\endgroup\@href {#1}{\urlprefix }}%
\providecommand \urlprefix  [0]{URL }%
\providecommand \Eprint [0]{\href }%
\providecommand \doibase [0]{http://dx.doi.org/}%
\providecommand \selectlanguage [0]{\@gobble}%
\providecommand \bibinfo  [0]{\@secondoftwo}%
\providecommand \bibfield  [0]{\@secondoftwo}%
\providecommand \translation [1]{[#1]}%
\providecommand \BibitemOpen [0]{}%
\providecommand \bibitemStop [0]{}%
\providecommand \bibitemNoStop [0]{.\EOS\space}%
\providecommand \EOS [0]{\spacefactor3000\relax}%
\providecommand \BibitemShut  [1]{\csname bibitem#1\endcsname}%
\let\auto@bib@innerbib\@empty
%</preamble>
\bibitem [{\citenamefont {Monroe}\ and\ \citenamefont
  {Kim}(2013)}]{Monroe2013}%
  \BibitemOpen
  \bibfield  {author} {\bibinfo {author} {\bibfnamefont {C.}~\bibnamefont
  {Monroe}}\ and\ \bibinfo {author} {\bibfnamefont {J.}~\bibnamefont {Kim}},\
  }\href@noop {} {\bibfield  {journal} {\bibinfo  {journal} {Science}\ }\textbf
  {\bibinfo {volume} {339}},\ \bibinfo {pages} {1164} (\bibinfo {year}
  {2013})}\BibitemShut {NoStop}%
\bibitem [{\citenamefont {Brown}\ \emph {et~al.}(2016)\citenamefont {Brown},
  \citenamefont {Kim},\ and\ \citenamefont {Monroe}}]{Brown2016}%
  \BibitemOpen
  \bibfield  {author} {\bibinfo {author} {\bibfnamefont {K.~R.}\ \bibnamefont
  {Brown}}, \bibinfo {author} {\bibfnamefont {J.}~\bibnamefont {Kim}}, \ and\
  \bibinfo {author} {\bibfnamefont {C.}~\bibnamefont {Monroe}},\ }\href@noop {}
  {\bibfield  {journal} {\bibinfo  {journal} {npj Quantum Information}\
  }\textbf {\bibinfo {volume} {2}},\ \bibinfo {pages} {1} (\bibinfo {year}
  {2016})}\BibitemShut {NoStop}%
\bibitem [{\citenamefont {Wang}\ \emph {et~al.}(2017)\citenamefont {Wang},
  \citenamefont {Um}, \citenamefont {Zhang}, \citenamefont {An}, \citenamefont
  {Lyu}, \citenamefont {Zhang}, \citenamefont {Duan}, \citenamefont {Yum},\
  and\ \citenamefont {Kim}}]{Wang2017}%
  \BibitemOpen
  \bibfield  {author} {\bibinfo {author} {\bibfnamefont {Y.}~\bibnamefont
  {Wang}}, \bibinfo {author} {\bibfnamefont {M.}~\bibnamefont {Um}}, \bibinfo
  {author} {\bibfnamefont {J.}~\bibnamefont {Zhang}}, \bibinfo {author}
  {\bibfnamefont {S.}~\bibnamefont {An}}, \bibinfo {author} {\bibfnamefont
  {M.}~\bibnamefont {Lyu}}, \bibinfo {author} {\bibfnamefont {J.-N.}\
  \bibnamefont {Zhang}}, \bibinfo {author} {\bibfnamefont {L.-M.}\ \bibnamefont
  {Duan}}, \bibinfo {author} {\bibfnamefont {D.}~\bibnamefont {Yum}}, \ and\
  \bibinfo {author} {\bibfnamefont {K.}~\bibnamefont {Kim}},\ }\href@noop {}
  {\bibfield  {journal} {\bibinfo  {journal} {Nature Photonics}\ }\textbf
  {\bibinfo {volume} {11}},\ \bibinfo {pages} {646} (\bibinfo {year}
  {2017})}\BibitemShut {NoStop}%
\bibitem [{\citenamefont {Wang}\ \emph {et~al.}(2021)\citenamefont {Wang},
  \citenamefont {Luan}, \citenamefont {Qiao}, \citenamefont {Um}, \citenamefont
  {Zhang}, \citenamefont {Wang}, \citenamefont {Yuan}, \citenamefont {Gu},
  \citenamefont {Zhang},\ and\ \citenamefont {Kim}}]{Wang2021}%
  \BibitemOpen
  \bibfield  {author} {\bibinfo {author} {\bibfnamefont {P.}~\bibnamefont
  {Wang}}, \bibinfo {author} {\bibfnamefont {C.-Y.}\ \bibnamefont {Luan}},
  \bibinfo {author} {\bibfnamefont {M.}~\bibnamefont {Qiao}}, \bibinfo {author}
  {\bibfnamefont {M.}~\bibnamefont {Um}}, \bibinfo {author} {\bibfnamefont
  {J.}~\bibnamefont {Zhang}}, \bibinfo {author} {\bibfnamefont
  {Y.}~\bibnamefont {Wang}}, \bibinfo {author} {\bibfnamefont {X.}~\bibnamefont
  {Yuan}}, \bibinfo {author} {\bibfnamefont {M.}~\bibnamefont {Gu}}, \bibinfo
  {author} {\bibfnamefont {J.}~\bibnamefont {Zhang}}, \ and\ \bibinfo {author}
  {\bibfnamefont {K.}~\bibnamefont {Kim}},\ }\href@noop {} {\bibfield
  {journal} {\bibinfo  {journal} {Nature communications}\ }\textbf {\bibinfo
  {volume} {12}},\ \bibinfo {pages} {1} (\bibinfo {year} {2021})}\BibitemShut
  {NoStop}%
\bibitem [{\citenamefont {Olmschenk}\ \emph {et~al.}(2007)\citenamefont
  {Olmschenk}, \citenamefont {Younge}, \citenamefont {Moehring}, \citenamefont
  {Matsukevich}, \citenamefont {Maunz},\ and\ \citenamefont
  {Monroe}}]{Olmschenk2007}%
  \BibitemOpen
  \bibfield  {author} {\bibinfo {author} {\bibfnamefont {S.}~\bibnamefont
  {Olmschenk}}, \bibinfo {author} {\bibfnamefont {K.~C.}\ \bibnamefont
  {Younge}}, \bibinfo {author} {\bibfnamefont {D.~L.}\ \bibnamefont
  {Moehring}}, \bibinfo {author} {\bibfnamefont {D.~N.}\ \bibnamefont
  {Matsukevich}}, \bibinfo {author} {\bibfnamefont {P.}~\bibnamefont {Maunz}},
  \ and\ \bibinfo {author} {\bibfnamefont {C.}~\bibnamefont {Monroe}},\ }\href
  {\doibase 10.1103/PhysRevA.76.052314} {\bibfield  {journal} {\bibinfo
  {journal} {Phys. Rev. A}\ }\textbf {\bibinfo {volume} {76}},\ \bibinfo
  {pages} {052314} (\bibinfo {year} {2007})}\BibitemShut {NoStop}%
\bibitem [{\citenamefont {Noek}\ \emph {et~al.}(2013)\citenamefont {Noek},
  \citenamefont {Vrijsen}, \citenamefont {Gaultney}, \citenamefont {Mount},
  \citenamefont {Kim}, \citenamefont {Maunz},\ and\ \citenamefont
  {Kim}}]{Noek2013}%
  \BibitemOpen
  \bibfield  {author} {\bibinfo {author} {\bibfnamefont {R.}~\bibnamefont
  {Noek}}, \bibinfo {author} {\bibfnamefont {G.}~\bibnamefont {Vrijsen}},
  \bibinfo {author} {\bibfnamefont {D.}~\bibnamefont {Gaultney}}, \bibinfo
  {author} {\bibfnamefont {E.}~\bibnamefont {Mount}}, \bibinfo {author}
  {\bibfnamefont {T.}~\bibnamefont {Kim}}, \bibinfo {author} {\bibfnamefont
  {P.}~\bibnamefont {Maunz}}, \ and\ \bibinfo {author} {\bibfnamefont
  {J.}~\bibnamefont {Kim}},\ }\href@noop {} {\bibfield  {journal} {\bibinfo
  {journal} {Opt. Lett.}\ }\textbf {\bibinfo {volume} {38}},\ \bibinfo {pages}
  {4735} (\bibinfo {year} {2013})}\BibitemShut {NoStop}%
\bibitem [{\citenamefont {Harty}\ \emph {et~al.}(2014)\citenamefont {Harty},
  \citenamefont {Allcock}, \citenamefont {Ballance}, \citenamefont {Guidoni},
  \citenamefont {Janacek}, \citenamefont {Linke}, \citenamefont {Stacey},\ and\
  \citenamefont {Lucas}}]{Harty2014}%
  \BibitemOpen
  \bibfield  {author} {\bibinfo {author} {\bibfnamefont {T.~P.}\ \bibnamefont
  {Harty}}, \bibinfo {author} {\bibfnamefont {D.~T.~C.}\ \bibnamefont
  {Allcock}}, \bibinfo {author} {\bibfnamefont {C.~J.}\ \bibnamefont
  {Ballance}}, \bibinfo {author} {\bibfnamefont {L.}~\bibnamefont {Guidoni}},
  \bibinfo {author} {\bibfnamefont {H.~A.}\ \bibnamefont {Janacek}}, \bibinfo
  {author} {\bibfnamefont {N.~M.}\ \bibnamefont {Linke}}, \bibinfo {author}
  {\bibfnamefont {D.~N.}\ \bibnamefont {Stacey}}, \ and\ \bibinfo {author}
  {\bibfnamefont {D.~M.}\ \bibnamefont {Lucas}},\ }\href {\doibase
  10.1103/PhysRevLett.113.220501} {\bibfield  {journal} {\bibinfo  {journal}
  {Phys. Rev. Lett.}\ }\textbf {\bibinfo {volume} {113}},\ \bibinfo {pages}
  {220501} (\bibinfo {year} {2014})}\BibitemShut {NoStop}%
\bibitem [{\citenamefont {Brown}\ \emph {et~al.}(2011)\citenamefont {Brown},
  \citenamefont {Wilson}, \citenamefont {Colombe}, \citenamefont {Ospelkaus},
  \citenamefont {Meier}, \citenamefont {Knill}, \citenamefont {Leibfried},\
  and\ \citenamefont {Wineland}}]{Brown2011}%
  \BibitemOpen
  \bibfield  {author} {\bibinfo {author} {\bibfnamefont {K.~R.}\ \bibnamefont
  {Brown}}, \bibinfo {author} {\bibfnamefont {A.~C.}\ \bibnamefont {Wilson}},
  \bibinfo {author} {\bibfnamefont {Y.}~\bibnamefont {Colombe}}, \bibinfo
  {author} {\bibfnamefont {C.}~\bibnamefont {Ospelkaus}}, \bibinfo {author}
  {\bibfnamefont {A.~M.}\ \bibnamefont {Meier}}, \bibinfo {author}
  {\bibfnamefont {E.}~\bibnamefont {Knill}}, \bibinfo {author} {\bibfnamefont
  {D.}~\bibnamefont {Leibfried}}, \ and\ \bibinfo {author} {\bibfnamefont
  {D.~J.}\ \bibnamefont {Wineland}},\ }\href {\doibase
  10.1103/PhysRevA.84.030303} {\bibfield  {journal} {\bibinfo  {journal} {Phys.
  Rev. A}\ }\textbf {\bibinfo {volume} {84}},\ \bibinfo {pages} {030303}
  (\bibinfo {year} {2011})}\BibitemShut {NoStop}%
\bibitem [{\citenamefont {Aude~Craik}\ \emph {et~al.}(2017)\citenamefont
  {Aude~Craik}, \citenamefont {Linke}, \citenamefont {Sepiol}, \citenamefont
  {Harty}, \citenamefont {Goodwin}, \citenamefont {Ballance}, \citenamefont
  {Stacey}, \citenamefont {Steane}, \citenamefont {Lucas},\ and\ \citenamefont
  {Allcock}}]{AudeCraik2017}%
  \BibitemOpen
  \bibfield  {author} {\bibinfo {author} {\bibfnamefont {D.~P.~L.}\
  \bibnamefont {Aude~Craik}}, \bibinfo {author} {\bibfnamefont {N.~M.}\
  \bibnamefont {Linke}}, \bibinfo {author} {\bibfnamefont {M.~A.}\ \bibnamefont
  {Sepiol}}, \bibinfo {author} {\bibfnamefont {T.~P.}\ \bibnamefont {Harty}},
  \bibinfo {author} {\bibfnamefont {J.~F.}\ \bibnamefont {Goodwin}}, \bibinfo
  {author} {\bibfnamefont {C.~J.}\ \bibnamefont {Ballance}}, \bibinfo {author}
  {\bibfnamefont {D.~N.}\ \bibnamefont {Stacey}}, \bibinfo {author}
  {\bibfnamefont {A.~M.}\ \bibnamefont {Steane}}, \bibinfo {author}
  {\bibfnamefont {D.~M.}\ \bibnamefont {Lucas}}, \ and\ \bibinfo {author}
  {\bibfnamefont {D.~T.~C.}\ \bibnamefont {Allcock}},\ }\href {\doibase
  10.1103/PhysRevA.95.022337} {\bibfield  {journal} {\bibinfo  {journal} {Phys.
  Rev. A}\ }\textbf {\bibinfo {volume} {95}},\ \bibinfo {pages} {022337}
  (\bibinfo {year} {2017})}\BibitemShut {NoStop}%
\bibitem [{\citenamefont {Ballance}\ \emph {et~al.}(2016)\citenamefont
  {Ballance}, \citenamefont {Harty}, \citenamefont {Linke}, \citenamefont
  {Sepiol},\ and\ \citenamefont {Lucas}}]{Ballance2016}%
  \BibitemOpen
  \bibfield  {author} {\bibinfo {author} {\bibfnamefont {C.~J.}\ \bibnamefont
  {Ballance}}, \bibinfo {author} {\bibfnamefont {T.~P.}\ \bibnamefont {Harty}},
  \bibinfo {author} {\bibfnamefont {N.~M.}\ \bibnamefont {Linke}}, \bibinfo
  {author} {\bibfnamefont {M.~A.}\ \bibnamefont {Sepiol}}, \ and\ \bibinfo
  {author} {\bibfnamefont {D.~M.}\ \bibnamefont {Lucas}},\ }\href {\doibase
  10.1103/PhysRevLett.117.060504} {\bibfield  {journal} {\bibinfo  {journal}
  {Phys. Rev. Lett.}\ }\textbf {\bibinfo {volume} {117}},\ \bibinfo {pages}
  {060504} (\bibinfo {year} {2016})}\BibitemShut {NoStop}%
\bibitem [{\citenamefont {Gaebler}\ \emph {et~al.}(2016)\citenamefont
  {Gaebler}, \citenamefont {Tan}, \citenamefont {Lin}, \citenamefont {Wan},
  \citenamefont {Bowler}, \citenamefont {Keith}, \citenamefont {Glancy},
  \citenamefont {Coakley}, \citenamefont {Knill}, \citenamefont {Leibfried},\
  and\ \citenamefont {Wineland}}]{Gaebler2016}%
  \BibitemOpen
  \bibfield  {author} {\bibinfo {author} {\bibfnamefont {J.~P.}\ \bibnamefont
  {Gaebler}}, \bibinfo {author} {\bibfnamefont {T.~R.}\ \bibnamefont {Tan}},
  \bibinfo {author} {\bibfnamefont {Y.}~\bibnamefont {Lin}}, \bibinfo {author}
  {\bibfnamefont {Y.}~\bibnamefont {Wan}}, \bibinfo {author} {\bibfnamefont
  {R.}~\bibnamefont {Bowler}}, \bibinfo {author} {\bibfnamefont {A.~C.}\
  \bibnamefont {Keith}}, \bibinfo {author} {\bibfnamefont {S.}~\bibnamefont
  {Glancy}}, \bibinfo {author} {\bibfnamefont {K.}~\bibnamefont {Coakley}},
  \bibinfo {author} {\bibfnamefont {E.}~\bibnamefont {Knill}}, \bibinfo
  {author} {\bibfnamefont {D.}~\bibnamefont {Leibfried}}, \ and\ \bibinfo
  {author} {\bibfnamefont {D.~J.}\ \bibnamefont {Wineland}},\ }\href {\doibase
  10.1103/PhysRevLett.117.060505} {\bibfield  {journal} {\bibinfo  {journal}
  {Phys. Rev. Lett.}\ }\textbf {\bibinfo {volume} {117}},\ \bibinfo {pages}
  {060505} (\bibinfo {year} {2016})}\BibitemShut {NoStop}%
\bibitem [{\citenamefont {Wang}\ \emph {et~al.}(2020)\citenamefont {Wang},
  \citenamefont {Crain}, \citenamefont {Fang}, \citenamefont {Zhang},
  \citenamefont {Huang}, \citenamefont {Liang}, \citenamefont {Leung},
  \citenamefont {Brown},\ and\ \citenamefont {Kim}}]{Wang2020}%
  \BibitemOpen
  \bibfield  {author} {\bibinfo {author} {\bibfnamefont {Y.}~\bibnamefont
  {Wang}}, \bibinfo {author} {\bibfnamefont {S.}~\bibnamefont {Crain}},
  \bibinfo {author} {\bibfnamefont {C.}~\bibnamefont {Fang}}, \bibinfo {author}
  {\bibfnamefont {B.}~\bibnamefont {Zhang}}, \bibinfo {author} {\bibfnamefont
  {S.}~\bibnamefont {Huang}}, \bibinfo {author} {\bibfnamefont
  {Q.}~\bibnamefont {Liang}}, \bibinfo {author} {\bibfnamefont {P.~H.}\
  \bibnamefont {Leung}}, \bibinfo {author} {\bibfnamefont {K.~R.}\ \bibnamefont
  {Brown}}, \ and\ \bibinfo {author} {\bibfnamefont {J.}~\bibnamefont {Kim}},\
  }\href {\doibase 10.1103/PhysRevLett.125.150505} {\bibfield  {journal}
  {\bibinfo  {journal} {Phys. Rev. Lett.}\ }\textbf {\bibinfo {volume} {125}},\
  \bibinfo {pages} {150505} (\bibinfo {year} {2020})}\BibitemShut {NoStop}%
\bibitem [{\citenamefont {Clark}\ \emph {et~al.}(2021)\citenamefont {Clark},
  \citenamefont {Tinkey}, \citenamefont {Sawyer}, \citenamefont {Meier},
  \citenamefont {Burkhardt}, \citenamefont {Seck}, \citenamefont {Shappert},
  \citenamefont {Guise}, \citenamefont {Volin}, \citenamefont {Fallek},
  \citenamefont {Hayden}, \citenamefont {Rellergert},\ and\ \citenamefont
  {Brown}}]{Clark2021}%
  \BibitemOpen
  \bibfield  {author} {\bibinfo {author} {\bibfnamefont {C.~R.}\ \bibnamefont
  {Clark}}, \bibinfo {author} {\bibfnamefont {H.~N.}\ \bibnamefont {Tinkey}},
  \bibinfo {author} {\bibfnamefont {B.~C.}\ \bibnamefont {Sawyer}}, \bibinfo
  {author} {\bibfnamefont {A.~M.}\ \bibnamefont {Meier}}, \bibinfo {author}
  {\bibfnamefont {K.~A.}\ \bibnamefont {Burkhardt}}, \bibinfo {author}
  {\bibfnamefont {C.~M.}\ \bibnamefont {Seck}}, \bibinfo {author}
  {\bibfnamefont {C.~M.}\ \bibnamefont {Shappert}}, \bibinfo {author}
  {\bibfnamefont {N.~D.}\ \bibnamefont {Guise}}, \bibinfo {author}
  {\bibfnamefont {C.~E.}\ \bibnamefont {Volin}}, \bibinfo {author}
  {\bibfnamefont {S.~D.}\ \bibnamefont {Fallek}}, \bibinfo {author}
  {\bibfnamefont {H.~T.}\ \bibnamefont {Hayden}}, \bibinfo {author}
  {\bibfnamefont {W.~G.}\ \bibnamefont {Rellergert}}, \ and\ \bibinfo {author}
  {\bibfnamefont {K.~R.}\ \bibnamefont {Brown}},\ }\href {\doibase
  10.1103/PhysRevLett.127.130505} {\bibfield  {journal} {\bibinfo  {journal}
  {Phys. Rev. Lett.}\ }\textbf {\bibinfo {volume} {127}},\ \bibinfo {pages}
  {130505} (\bibinfo {year} {2021})}\BibitemShut {NoStop}%
\bibitem [{\citenamefont {Baldwin}\ \emph {et~al.}(2021)\citenamefont
  {Baldwin}, \citenamefont {Bjork}, \citenamefont {Foss-Feig}, \citenamefont
  {Gaebler}, \citenamefont {Hayes}, \citenamefont {Kokish}, \citenamefont
  {Langer}, \citenamefont {Sedlacek}, \citenamefont {Stack},\ and\
  \citenamefont {Vittorini}}]{Baldwin2021}%
  \BibitemOpen
  \bibfield  {author} {\bibinfo {author} {\bibfnamefont {C.~H.}\ \bibnamefont
  {Baldwin}}, \bibinfo {author} {\bibfnamefont {B.~J.}\ \bibnamefont {Bjork}},
  \bibinfo {author} {\bibfnamefont {M.}~\bibnamefont {Foss-Feig}}, \bibinfo
  {author} {\bibfnamefont {J.~P.}\ \bibnamefont {Gaebler}}, \bibinfo {author}
  {\bibfnamefont {D.}~\bibnamefont {Hayes}}, \bibinfo {author} {\bibfnamefont
  {M.~G.}\ \bibnamefont {Kokish}}, \bibinfo {author} {\bibfnamefont
  {C.}~\bibnamefont {Langer}}, \bibinfo {author} {\bibfnamefont {J.~A.}\
  \bibnamefont {Sedlacek}}, \bibinfo {author} {\bibfnamefont {D.}~\bibnamefont
  {Stack}}, \ and\ \bibinfo {author} {\bibfnamefont {G.}~\bibnamefont
  {Vittorini}},\ }\href {\doibase 10.1103/PhysRevA.103.012603} {\bibfield
  {journal} {\bibinfo  {journal} {Phys. Rev. A}\ }\textbf {\bibinfo {volume}
  {103}},\ \bibinfo {pages} {012603} (\bibinfo {year} {2021})}\BibitemShut
  {NoStop}%
\bibitem [{\citenamefont {Srinivas}\ \emph {et~al.}(2021)\citenamefont
  {Srinivas}, \citenamefont {Burd}, \citenamefont {Knaack}, \citenamefont
  {Sutherland}, \citenamefont {Kwiatkowski}, \citenamefont {Glancy},
  \citenamefont {Knill}, \citenamefont {Wineland}, \citenamefont {Leibfried},
  \citenamefont {Wilson} \emph {et~al.}}]{Srinivas2021}%
  \BibitemOpen
  \bibfield  {author} {\bibinfo {author} {\bibfnamefont {R.}~\bibnamefont
  {Srinivas}}, \bibinfo {author} {\bibfnamefont {S.}~\bibnamefont {Burd}},
  \bibinfo {author} {\bibfnamefont {H.}~\bibnamefont {Knaack}}, \bibinfo
  {author} {\bibfnamefont {R.}~\bibnamefont {Sutherland}}, \bibinfo {author}
  {\bibfnamefont {A.}~\bibnamefont {Kwiatkowski}}, \bibinfo {author}
  {\bibfnamefont {S.}~\bibnamefont {Glancy}}, \bibinfo {author} {\bibfnamefont
  {E.}~\bibnamefont {Knill}}, \bibinfo {author} {\bibfnamefont
  {D.}~\bibnamefont {Wineland}}, \bibinfo {author} {\bibfnamefont
  {D.}~\bibnamefont {Leibfried}}, \bibinfo {author} {\bibfnamefont {A.~C.}\
  \bibnamefont {Wilson}},  \emph {et~al.},\ }\href@noop {} {\bibfield
  {journal} {\bibinfo  {journal} {Nature (London)}\ }\textbf {\bibinfo {volume}
  {597}},\ \bibinfo {pages} {209} (\bibinfo {year} {2021})}\BibitemShut
  {NoStop}%
\bibitem [{\citenamefont {M\o{}lmer}\ and\ \citenamefont
  {S\o{}rensen}(1999)}]{Molmer1999}%
  \BibitemOpen
  \bibfield  {author} {\bibinfo {author} {\bibfnamefont {K.}~\bibnamefont
  {M\o{}lmer}}\ and\ \bibinfo {author} {\bibfnamefont {A.}~\bibnamefont
  {S\o{}rensen}},\ }\href {\doibase 10.1103/PhysRevLett.82.1835} {\bibfield
  {journal} {\bibinfo  {journal} {Phys. Rev. Lett.}\ }\textbf {\bibinfo
  {volume} {82}},\ \bibinfo {pages} {1835} (\bibinfo {year}
  {1999})}\BibitemShut {NoStop}%
\bibitem [{\citenamefont {S\o{}rensen}\ and\ \citenamefont
  {M\o{}lmer}(2000)}]{Sorensen2000}%
  \BibitemOpen
  \bibfield  {author} {\bibinfo {author} {\bibfnamefont {A.}~\bibnamefont
  {S\o{}rensen}}\ and\ \bibinfo {author} {\bibfnamefont {K.}~\bibnamefont
  {M\o{}lmer}},\ }\href {\doibase 10.1103/PhysRevA.62.022311} {\bibfield
  {journal} {\bibinfo  {journal} {Phys. Rev. A}\ }\textbf {\bibinfo {volume}
  {62}},\ \bibinfo {pages} {022311} (\bibinfo {year} {2000})}\BibitemShut
  {NoStop}%
\bibitem [{\citenamefont {Roos}(2008)}]{Roos2008}%
  \BibitemOpen
  \bibfield  {author} {\bibinfo {author} {\bibfnamefont {C.~F.}\ \bibnamefont
  {Roos}},\ }\href {\doibase 10.1088/1367-2630/10/1/013002} {\bibfield
  {journal} {\bibinfo  {journal} {New Journal of Physics}\ }\textbf {\bibinfo
  {volume} {10}},\ \bibinfo {pages} {013002} (\bibinfo {year}
  {2008})}\BibitemShut {NoStop}%
\bibitem [{\citenamefont {Debnath}\ \emph {et~al.}(2016)\citenamefont
  {Debnath}, \citenamefont {Linke}, \citenamefont {Figgatt}, \citenamefont
  {Landsman}, \citenamefont {Wright},\ and\ \citenamefont
  {Monroe}}]{Debnath2016}%
  \BibitemOpen
  \bibfield  {author} {\bibinfo {author} {\bibfnamefont {S.}~\bibnamefont
  {Debnath}}, \bibinfo {author} {\bibfnamefont {N.~M.}\ \bibnamefont {Linke}},
  \bibinfo {author} {\bibfnamefont {C.}~\bibnamefont {Figgatt}}, \bibinfo
  {author} {\bibfnamefont {K.~A.}\ \bibnamefont {Landsman}}, \bibinfo {author}
  {\bibfnamefont {K.}~\bibnamefont {Wright}}, \ and\ \bibinfo {author}
  {\bibfnamefont {C.}~\bibnamefont {Monroe}},\ }\href@noop {} {\bibfield
  {journal} {\bibinfo  {journal} {Nature}\ }\textbf {\bibinfo {volume} {536}},\
  \bibinfo {pages} {63} (\bibinfo {year} {2016})}\BibitemShut {NoStop}%
\bibitem [{\citenamefont {Wu}\ \emph {et~al.}(2018)\citenamefont {Wu},
  \citenamefont {Wang},\ and\ \citenamefont {Duan}}]{Wu2018}%
  \BibitemOpen
  \bibfield  {author} {\bibinfo {author} {\bibfnamefont {Y.}~\bibnamefont
  {Wu}}, \bibinfo {author} {\bibfnamefont {S.-T.}\ \bibnamefont {Wang}}, \ and\
  \bibinfo {author} {\bibfnamefont {L.-M.}\ \bibnamefont {Duan}},\ }\href
  {\doibase 10.1103/PhysRevA.97.062325} {\bibfield  {journal} {\bibinfo
  {journal} {Phys. Rev. A}\ }\textbf {\bibinfo {volume} {97}},\ \bibinfo
  {pages} {062325} (\bibinfo {year} {2018})}\BibitemShut {NoStop}%
\bibitem [{\citenamefont {Figgatt}\ \emph {et~al.}(2019)\citenamefont
  {Figgatt}, \citenamefont {Ostrander}, \citenamefont {Linke}, \citenamefont
  {Landsman}, \citenamefont {Zhu}, \citenamefont {Maslov},\ and\ \citenamefont
  {Monroe}}]{Figgatt2019}%
  \BibitemOpen
  \bibfield  {author} {\bibinfo {author} {\bibfnamefont {C.}~\bibnamefont
  {Figgatt}}, \bibinfo {author} {\bibfnamefont {A.}~\bibnamefont {Ostrander}},
  \bibinfo {author} {\bibfnamefont {N.~M.}\ \bibnamefont {Linke}}, \bibinfo
  {author} {\bibfnamefont {K.~A.}\ \bibnamefont {Landsman}}, \bibinfo {author}
  {\bibfnamefont {D.}~\bibnamefont {Zhu}}, \bibinfo {author} {\bibfnamefont
  {D.}~\bibnamefont {Maslov}}, \ and\ \bibinfo {author} {\bibfnamefont
  {C.}~\bibnamefont {Monroe}},\ }\href@noop {} {\bibfield  {journal} {\bibinfo
  {journal} {Nature}\ }\textbf {\bibinfo {volume} {572}},\ \bibinfo {pages}
  {368} (\bibinfo {year} {2019})}\BibitemShut {NoStop}%
\bibitem [{\citenamefont {Bl{\"u}mel}\ \emph {et~al.}(2021)\citenamefont
  {Bl{\"u}mel}, \citenamefont {Grzesiak}, \citenamefont {Pisenti},
  \citenamefont {Wright},\ and\ \citenamefont {Nam}}]{blumel2021}%
  \BibitemOpen
  \bibfield  {author} {\bibinfo {author} {\bibfnamefont {R.}~\bibnamefont
  {Bl{\"u}mel}}, \bibinfo {author} {\bibfnamefont {N.}~\bibnamefont
  {Grzesiak}}, \bibinfo {author} {\bibfnamefont {N.}~\bibnamefont {Pisenti}},
  \bibinfo {author} {\bibfnamefont {K.}~\bibnamefont {Wright}}, \ and\ \bibinfo
  {author} {\bibfnamefont {Y.}~\bibnamefont {Nam}},\ }\href
  {https://www.nature.com/articles/s41534-021-00489-w} {\bibfield  {journal}
  {\bibinfo  {journal} {npj Quantum Information}\ }\textbf {\bibinfo {volume}
  {7}},\ \bibinfo {pages} {1} (\bibinfo {year} {2021})}\BibitemShut {NoStop}%
\bibitem [{\citenamefont {Cetina}\ \emph {et~al.}(2022)\citenamefont {Cetina},
  \citenamefont {Egan}, \citenamefont {Noel}, \citenamefont {Goldman},
  \citenamefont {Biswas}, \citenamefont {Risinger}, \citenamefont {Zhu},\ and\
  \citenamefont {Monroe}}]{Cetina2022}%
  \BibitemOpen
  \bibfield  {author} {\bibinfo {author} {\bibfnamefont {M.}~\bibnamefont
  {Cetina}}, \bibinfo {author} {\bibfnamefont {L.}~\bibnamefont {Egan}},
  \bibinfo {author} {\bibfnamefont {C.}~\bibnamefont {Noel}}, \bibinfo {author}
  {\bibfnamefont {M.}~\bibnamefont {Goldman}}, \bibinfo {author} {\bibfnamefont
  {D.}~\bibnamefont {Biswas}}, \bibinfo {author} {\bibfnamefont
  {A.}~\bibnamefont {Risinger}}, \bibinfo {author} {\bibfnamefont
  {D.}~\bibnamefont {Zhu}}, \ and\ \bibinfo {author} {\bibfnamefont
  {C.}~\bibnamefont {Monroe}},\ }\href {\doibase 10.1103/PRXQuantum.3.010334}
  {\bibfield  {journal} {\bibinfo  {journal} {PRX Quantum}\ }\textbf {\bibinfo
  {volume} {3}},\ \bibinfo {pages} {010334} (\bibinfo {year}
  {2022})}\BibitemShut {NoStop}%
\bibitem [{\citenamefont {Leung}\ \emph {et~al.}(2018)\citenamefont {Leung},
  \citenamefont {Landsman}, \citenamefont {Figgatt}, \citenamefont {Linke},
  \citenamefont {Monroe},\ and\ \citenamefont {Brown}}]{Leung2018}%
  \BibitemOpen
  \bibfield  {author} {\bibinfo {author} {\bibfnamefont {P.~H.}\ \bibnamefont
  {Leung}}, \bibinfo {author} {\bibfnamefont {K.~A.}\ \bibnamefont {Landsman}},
  \bibinfo {author} {\bibfnamefont {C.}~\bibnamefont {Figgatt}}, \bibinfo
  {author} {\bibfnamefont {N.~M.}\ \bibnamefont {Linke}}, \bibinfo {author}
  {\bibfnamefont {C.}~\bibnamefont {Monroe}}, \ and\ \bibinfo {author}
  {\bibfnamefont {K.~R.}\ \bibnamefont {Brown}},\ }\href {\doibase
  10.1103/PhysRevLett.120.020501} {\bibfield  {journal} {\bibinfo  {journal}
  {Phys. Rev. Lett.}\ }\textbf {\bibinfo {volume} {120}},\ \bibinfo {pages}
  {020501} (\bibinfo {year} {2018})}\BibitemShut {NoStop}%
\bibitem [{\citenamefont {Kang}\ \emph {et~al.}(2022)\citenamefont {Kang},
  \citenamefont {Wang}, \citenamefont {Fang}, \citenamefont {Zhang},
  \citenamefont {Khosravani}, \citenamefont {Kim},\ and\ \citenamefont
  {Brown}}]{Kang2022}%
  \BibitemOpen
  \bibfield  {author} {\bibinfo {author} {\bibfnamefont {M.}~\bibnamefont
  {Kang}}, \bibinfo {author} {\bibfnamefont {Y.}~\bibnamefont {Wang}}, \bibinfo
  {author} {\bibfnamefont {C.}~\bibnamefont {Fang}}, \bibinfo {author}
  {\bibfnamefont {B.}~\bibnamefont {Zhang}}, \bibinfo {author} {\bibfnamefont
  {O.}~\bibnamefont {Khosravani}}, \bibinfo {author} {\bibfnamefont
  {J.}~\bibnamefont {Kim}}, \ and\ \bibinfo {author} {\bibfnamefont {K.~R.}\
  \bibnamefont {Brown}},\ }\href@noop {} {\bibfield  {journal} {\bibinfo
  {journal} {arXiv preprint arXiv:2206.10850}\ } (\bibinfo {year}
  {2022})}\BibitemShut {NoStop}%
\bibitem [{\citenamefont {Fang}\ \emph {et~al.}(2022)\citenamefont {Fang},
  \citenamefont {Wang}, \citenamefont {Huang}, \citenamefont {Brown},\ and\
  \citenamefont {Kim}}]{Fang2022}%
  \BibitemOpen
  \bibfield  {author} {\bibinfo {author} {\bibfnamefont {C.}~\bibnamefont
  {Fang}}, \bibinfo {author} {\bibfnamefont {Y.}~\bibnamefont {Wang}}, \bibinfo
  {author} {\bibfnamefont {S.}~\bibnamefont {Huang}}, \bibinfo {author}
  {\bibfnamefont {K.~R.}\ \bibnamefont {Brown}}, \ and\ \bibinfo {author}
  {\bibfnamefont {J.}~\bibnamefont {Kim}},\ }\href@noop {} {\bibfield
  {journal} {\bibinfo  {journal} {arXiv preprint arXiv:2206.02703}\ } (\bibinfo
  {year} {2022})}\BibitemShut {NoStop}%
\bibitem [{\citenamefont {Green}\ and\ \citenamefont
  {Biercuk}(2015)}]{Green2015}%
  \BibitemOpen
  \bibfield  {author} {\bibinfo {author} {\bibfnamefont {T.~J.}\ \bibnamefont
  {Green}}\ and\ \bibinfo {author} {\bibfnamefont {M.~J.}\ \bibnamefont
  {Biercuk}},\ }\href {\doibase 10.1103/PhysRevLett.114.120502} {\bibfield
  {journal} {\bibinfo  {journal} {Phys. Rev. Lett.}\ }\textbf {\bibinfo
  {volume} {114}},\ \bibinfo {pages} {120502} (\bibinfo {year}
  {2015})}\BibitemShut {NoStop}%
\bibitem [{\citenamefont {Lu}\ \emph {et~al.}(2019)\citenamefont {Lu},
  \citenamefont {Zhang}, \citenamefont {Zhang}, \citenamefont {Chen},
  \citenamefont {Shen}, \citenamefont {Zhang}, \citenamefont {Zhang},\ and\
  \citenamefont {Kim}}]{Lu2019}%
  \BibitemOpen
  \bibfield  {author} {\bibinfo {author} {\bibfnamefont {Y.}~\bibnamefont
  {Lu}}, \bibinfo {author} {\bibfnamefont {S.}~\bibnamefont {Zhang}}, \bibinfo
  {author} {\bibfnamefont {K.}~\bibnamefont {Zhang}}, \bibinfo {author}
  {\bibfnamefont {W.}~\bibnamefont {Chen}}, \bibinfo {author} {\bibfnamefont
  {Y.}~\bibnamefont {Shen}}, \bibinfo {author} {\bibfnamefont {J.}~\bibnamefont
  {Zhang}}, \bibinfo {author} {\bibfnamefont {J.-N.}\ \bibnamefont {Zhang}}, \
  and\ \bibinfo {author} {\bibfnamefont {K.}~\bibnamefont {Kim}},\ }\href@noop
  {} {\bibfield  {journal} {\bibinfo  {journal} {Nature}\ }\textbf {\bibinfo
  {volume} {572}},\ \bibinfo {pages} {363} (\bibinfo {year}
  {2019})}\BibitemShut {NoStop}%
\bibitem [{\citenamefont {Milne}\ \emph {et~al.}(2020)\citenamefont {Milne},
  \citenamefont {Edmunds}, \citenamefont {Hempel}, \citenamefont {Roy},
  \citenamefont {Mavadia},\ and\ \citenamefont {Biercuk}}]{Milne2020}%
  \BibitemOpen
  \bibfield  {author} {\bibinfo {author} {\bibfnamefont {A.~R.}\ \bibnamefont
  {Milne}}, \bibinfo {author} {\bibfnamefont {C.~L.}\ \bibnamefont {Edmunds}},
  \bibinfo {author} {\bibfnamefont {C.}~\bibnamefont {Hempel}}, \bibinfo
  {author} {\bibfnamefont {F.}~\bibnamefont {Roy}}, \bibinfo {author}
  {\bibfnamefont {S.}~\bibnamefont {Mavadia}}, \ and\ \bibinfo {author}
  {\bibfnamefont {M.~J.}\ \bibnamefont {Biercuk}},\ }\href {\doibase
  10.1103/PhysRevApplied.13.024022} {\bibfield  {journal} {\bibinfo  {journal}
  {Phys. Rev. Applied}\ }\textbf {\bibinfo {volume} {13}},\ \bibinfo {pages}
  {024022} (\bibinfo {year} {2020})}\BibitemShut {NoStop}%
\bibitem [{\citenamefont {Haddadfarshi}\ and\ \citenamefont
  {Mintert}(2016)}]{Haddadfarshi2016}%
  \BibitemOpen
  \bibfield  {author} {\bibinfo {author} {\bibfnamefont {F.}~\bibnamefont
  {Haddadfarshi}}\ and\ \bibinfo {author} {\bibfnamefont {F.}~\bibnamefont
  {Mintert}},\ }\href {\doibase 10.1088/1367-2630/18/12/123007} {\bibfield
  {journal} {\bibinfo  {journal} {New Journal of Physics}\ }\textbf {\bibinfo
  {volume} {18}},\ \bibinfo {pages} {123007} (\bibinfo {year}
  {2016})}\BibitemShut {NoStop}%
\bibitem [{\citenamefont {Shapira}\ \emph {et~al.}(2018)\citenamefont
  {Shapira}, \citenamefont {Shaniv}, \citenamefont {Manovitz}, \citenamefont
  {Akerman},\ and\ \citenamefont {Ozeri}}]{Shapira2018}%
  \BibitemOpen
  \bibfield  {author} {\bibinfo {author} {\bibfnamefont {Y.}~\bibnamefont
  {Shapira}}, \bibinfo {author} {\bibfnamefont {R.}~\bibnamefont {Shaniv}},
  \bibinfo {author} {\bibfnamefont {T.}~\bibnamefont {Manovitz}}, \bibinfo
  {author} {\bibfnamefont {N.}~\bibnamefont {Akerman}}, \ and\ \bibinfo
  {author} {\bibfnamefont {R.}~\bibnamefont {Ozeri}},\ }\href {\doibase
  10.1103/PhysRevLett.121.180502} {\bibfield  {journal} {\bibinfo  {journal}
  {Phys. Rev. Lett.}\ }\textbf {\bibinfo {volume} {121}},\ \bibinfo {pages}
  {180502} (\bibinfo {year} {2018})}\BibitemShut {NoStop}%
\bibitem [{\citenamefont {Shapira}\ \emph {et~al.}(2020)\citenamefont
  {Shapira}, \citenamefont {Shaniv}, \citenamefont {Manovitz}, \citenamefont
  {Akerman}, \citenamefont {Peleg}, \citenamefont {Gazit}, \citenamefont
  {Ozeri},\ and\ \citenamefont {Stern}}]{Shapira2020}%
  \BibitemOpen
  \bibfield  {author} {\bibinfo {author} {\bibfnamefont {Y.}~\bibnamefont
  {Shapira}}, \bibinfo {author} {\bibfnamefont {R.}~\bibnamefont {Shaniv}},
  \bibinfo {author} {\bibfnamefont {T.}~\bibnamefont {Manovitz}}, \bibinfo
  {author} {\bibfnamefont {N.}~\bibnamefont {Akerman}}, \bibinfo {author}
  {\bibfnamefont {L.}~\bibnamefont {Peleg}}, \bibinfo {author} {\bibfnamefont
  {L.}~\bibnamefont {Gazit}}, \bibinfo {author} {\bibfnamefont
  {R.}~\bibnamefont {Ozeri}}, \ and\ \bibinfo {author} {\bibfnamefont
  {A.}~\bibnamefont {Stern}},\ }\href {\doibase 10.1103/PhysRevA.101.032330}
  {\bibfield  {journal} {\bibinfo  {journal} {Phys. Rev. A}\ }\textbf {\bibinfo
  {volume} {101}},\ \bibinfo {pages} {032330} (\bibinfo {year}
  {2020})}\BibitemShut {NoStop}%
\bibitem [{\citenamefont {Egan}\ \emph {et~al.}(2021)\citenamefont {Egan},
  \citenamefont {Debroy}, \citenamefont {Noel}, \citenamefont {Risinger},
  \citenamefont {Zhu}, \citenamefont {Biswas}, \citenamefont {Newman},
  \citenamefont {Li}, \citenamefont {Brown}, \citenamefont {Cetina} \emph
  {et~al.}}]{Egan2021}%
  \BibitemOpen
  \bibfield  {author} {\bibinfo {author} {\bibfnamefont {L.}~\bibnamefont
  {Egan}}, \bibinfo {author} {\bibfnamefont {D.~M.}\ \bibnamefont {Debroy}},
  \bibinfo {author} {\bibfnamefont {C.}~\bibnamefont {Noel}}, \bibinfo {author}
  {\bibfnamefont {A.}~\bibnamefont {Risinger}}, \bibinfo {author}
  {\bibfnamefont {D.}~\bibnamefont {Zhu}}, \bibinfo {author} {\bibfnamefont
  {D.}~\bibnamefont {Biswas}}, \bibinfo {author} {\bibfnamefont
  {M.}~\bibnamefont {Newman}}, \bibinfo {author} {\bibfnamefont
  {M.}~\bibnamefont {Li}}, \bibinfo {author} {\bibfnamefont {K.~R.}\
  \bibnamefont {Brown}}, \bibinfo {author} {\bibfnamefont {M.}~\bibnamefont
  {Cetina}},  \emph {et~al.},\ }\href@noop {} {\bibfield  {journal} {\bibinfo
  {journal} {Nature}\ }\textbf {\bibinfo {volume} {598}},\ \bibinfo {pages}
  {281} (\bibinfo {year} {2021})}\BibitemShut {NoStop}%
\bibitem [{\citenamefont {Wright}\ \emph {et~al.}(2019)\citenamefont {Wright},
  \citenamefont {Beck}, \citenamefont {Debnath}, \citenamefont {Amini},
  \citenamefont {Nam}, \citenamefont {Grzesiak}, \citenamefont {Chen},
  \citenamefont {Pisenti}, \citenamefont {Chmielewski}, \citenamefont {Collins}
  \emph {et~al.}}]{Wright2019}%
  \BibitemOpen
  \bibfield  {author} {\bibinfo {author} {\bibfnamefont {K.}~\bibnamefont
  {Wright}}, \bibinfo {author} {\bibfnamefont {K.~M.}\ \bibnamefont {Beck}},
  \bibinfo {author} {\bibfnamefont {S.}~\bibnamefont {Debnath}}, \bibinfo
  {author} {\bibfnamefont {J.}~\bibnamefont {Amini}}, \bibinfo {author}
  {\bibfnamefont {Y.}~\bibnamefont {Nam}}, \bibinfo {author} {\bibfnamefont
  {N.}~\bibnamefont {Grzesiak}}, \bibinfo {author} {\bibfnamefont {J.-S.}\
  \bibnamefont {Chen}}, \bibinfo {author} {\bibfnamefont {N.}~\bibnamefont
  {Pisenti}}, \bibinfo {author} {\bibfnamefont {M.}~\bibnamefont
  {Chmielewski}}, \bibinfo {author} {\bibfnamefont {C.}~\bibnamefont
  {Collins}},  \emph {et~al.},\ }\href@noop {} {\bibfield  {journal} {\bibinfo
  {journal} {Nature communications}\ }\textbf {\bibinfo {volume} {10}},\
  \bibinfo {pages} {1} (\bibinfo {year} {2019})}\BibitemShut {NoStop}%
\bibitem [{\citenamefont {Postler}\ \emph {et~al.}(2022)\citenamefont
  {Postler}, \citenamefont {Heu$\beta$en}, \citenamefont {Pogorelov},
  \citenamefont {Rispler}, \citenamefont {Feldker}, \citenamefont {Meth},
  \citenamefont {Marciniak}, \citenamefont {Stricker}, \citenamefont
  {Ringbauer}, \citenamefont {Blatt} \emph {et~al.}}]{Postler2022}%
  \BibitemOpen
  \bibfield  {author} {\bibinfo {author} {\bibfnamefont {L.}~\bibnamefont
  {Postler}}, \bibinfo {author} {\bibfnamefont {S.}~\bibnamefont
  {Heu$\beta$en}}, \bibinfo {author} {\bibfnamefont {I.}~\bibnamefont
  {Pogorelov}}, \bibinfo {author} {\bibfnamefont {M.}~\bibnamefont {Rispler}},
  \bibinfo {author} {\bibfnamefont {T.}~\bibnamefont {Feldker}}, \bibinfo
  {author} {\bibfnamefont {M.}~\bibnamefont {Meth}}, \bibinfo {author}
  {\bibfnamefont {C.~D.}\ \bibnamefont {Marciniak}}, \bibinfo {author}
  {\bibfnamefont {R.}~\bibnamefont {Stricker}}, \bibinfo {author}
  {\bibfnamefont {M.}~\bibnamefont {Ringbauer}}, \bibinfo {author}
  {\bibfnamefont {R.}~\bibnamefont {Blatt}},  \emph {et~al.},\ }\href@noop {}
  {\bibfield  {journal} {\bibinfo  {journal} {Nature}\ }\textbf {\bibinfo
  {volume} {605}},\ \bibinfo {pages} {675} (\bibinfo {year}
  {2022})}\BibitemShut {NoStop}%
\bibitem [{\citenamefont {Leung}\ and\ \citenamefont
  {Brown}(2018)}]{Leung2018LongChain}%
  \BibitemOpen
  \bibfield  {author} {\bibinfo {author} {\bibfnamefont {P.~H.}\ \bibnamefont
  {Leung}}\ and\ \bibinfo {author} {\bibfnamefont {K.~R.}\ \bibnamefont
  {Brown}},\ }\href {\doibase 10.1103/PhysRevA.98.032318} {\bibfield  {journal}
  {\bibinfo  {journal} {Phys. Rev. A}\ }\textbf {\bibinfo {volume} {98}},\
  \bibinfo {pages} {032318} (\bibinfo {year} {2018})}\BibitemShut {NoStop}%
\bibitem [{\citenamefont {Debroy}\ \emph {et~al.}(2018)\citenamefont {Debroy},
  \citenamefont {Li}, \citenamefont {Newman},\ and\ \citenamefont
  {Brown}}]{Debroy2018}%
  \BibitemOpen
  \bibfield  {author} {\bibinfo {author} {\bibfnamefont {D.~M.}\ \bibnamefont
  {Debroy}}, \bibinfo {author} {\bibfnamefont {M.}~\bibnamefont {Li}}, \bibinfo
  {author} {\bibfnamefont {M.}~\bibnamefont {Newman}}, \ and\ \bibinfo {author}
  {\bibfnamefont {K.~R.}\ \bibnamefont {Brown}},\ }\href {\doibase
  10.1103/PhysRevLett.121.250502} {\bibfield  {journal} {\bibinfo  {journal}
  {Phys. Rev. Lett.}\ }\textbf {\bibinfo {volume} {121}},\ \bibinfo {pages}
  {250502} (\bibinfo {year} {2018})}\BibitemShut {NoStop}%
\bibitem [{\citenamefont {Ryan-Anderson}\ \emph {et~al.}(2021)\citenamefont
  {Ryan-Anderson}, \citenamefont {Bohnet}, \citenamefont {Lee}, \citenamefont
  {Gresh}, \citenamefont {Hankin}, \citenamefont {Gaebler}, \citenamefont
  {Francois}, \citenamefont {Chernoguzov}, \citenamefont {Lucchetti},
  \citenamefont {Brown}, \citenamefont {Gatterman}, \citenamefont {Halit},
  \citenamefont {Gilmore}, \citenamefont {Gerber}, \citenamefont {Neyenhuis},
  \citenamefont {Hayes},\ and\ \citenamefont {Stutz}}]{RyanAnderson2021}%
  \BibitemOpen
  \bibfield  {author} {\bibinfo {author} {\bibfnamefont {C.}~\bibnamefont
  {Ryan-Anderson}}, \bibinfo {author} {\bibfnamefont {J.~G.}\ \bibnamefont
  {Bohnet}}, \bibinfo {author} {\bibfnamefont {K.}~\bibnamefont {Lee}},
  \bibinfo {author} {\bibfnamefont {D.}~\bibnamefont {Gresh}}, \bibinfo
  {author} {\bibfnamefont {A.}~\bibnamefont {Hankin}}, \bibinfo {author}
  {\bibfnamefont {J.~P.}\ \bibnamefont {Gaebler}}, \bibinfo {author}
  {\bibfnamefont {D.}~\bibnamefont {Francois}}, \bibinfo {author}
  {\bibfnamefont {A.}~\bibnamefont {Chernoguzov}}, \bibinfo {author}
  {\bibfnamefont {D.}~\bibnamefont {Lucchetti}}, \bibinfo {author}
  {\bibfnamefont {N.~C.}\ \bibnamefont {Brown}}, \bibinfo {author}
  {\bibfnamefont {T.~M.}\ \bibnamefont {Gatterman}}, \bibinfo {author}
  {\bibfnamefont {S.~K.}\ \bibnamefont {Halit}}, \bibinfo {author}
  {\bibfnamefont {K.}~\bibnamefont {Gilmore}}, \bibinfo {author} {\bibfnamefont
  {J.~A.}\ \bibnamefont {Gerber}}, \bibinfo {author} {\bibfnamefont
  {B.}~\bibnamefont {Neyenhuis}}, \bibinfo {author} {\bibfnamefont
  {D.}~\bibnamefont {Hayes}}, \ and\ \bibinfo {author} {\bibfnamefont {R.~P.}\
  \bibnamefont {Stutz}},\ }\href {\doibase 10.1103/PhysRevX.11.041058}
  {\bibfield  {journal} {\bibinfo  {journal} {Phys. Rev. X}\ }\textbf {\bibinfo
  {volume} {11}},\ \bibinfo {pages} {041058} (\bibinfo {year}
  {2021})}\BibitemShut {NoStop}%
\bibitem [{\citenamefont {Zhang}\ \emph {et~al.}(2022)\citenamefont {Zhang},
  \citenamefont {Majumder}, \citenamefont {Leung}, \citenamefont {Crain},
  \citenamefont {Wang}, \citenamefont {Fang}, \citenamefont {Debroy},
  \citenamefont {Kim},\ and\ \citenamefont {Brown}}]{Zhang2022}%
  \BibitemOpen
  \bibfield  {author} {\bibinfo {author} {\bibfnamefont {B.}~\bibnamefont
  {Zhang}}, \bibinfo {author} {\bibfnamefont {S.}~\bibnamefont {Majumder}},
  \bibinfo {author} {\bibfnamefont {P.~H.}\ \bibnamefont {Leung}}, \bibinfo
  {author} {\bibfnamefont {S.}~\bibnamefont {Crain}}, \bibinfo {author}
  {\bibfnamefont {Y.}~\bibnamefont {Wang}}, \bibinfo {author} {\bibfnamefont
  {C.}~\bibnamefont {Fang}}, \bibinfo {author} {\bibfnamefont {D.~M.}\
  \bibnamefont {Debroy}}, \bibinfo {author} {\bibfnamefont {J.}~\bibnamefont
  {Kim}}, \ and\ \bibinfo {author} {\bibfnamefont {K.~R.}\ \bibnamefont
  {Brown}},\ }\href {\doibase 10.1103/PhysRevApplied.17.034074} {\bibfield
  {journal} {\bibinfo  {journal} {Phys. Rev. Applied}\ }\textbf {\bibinfo
  {volume} {17}},\ \bibinfo {pages} {034074} (\bibinfo {year}
  {2022})}\BibitemShut {NoStop}%
\bibitem [{\citenamefont {Majumder}\ \emph {et~al.}(2022)\citenamefont
  {Majumder}, \citenamefont {Yale}, \citenamefont {Morris}, \citenamefont
  {Lobser}, \citenamefont {Burch}, \citenamefont {Chow}, \citenamefont
  {Revelle}, \citenamefont {Clark},\ and\ \citenamefont
  {Pooser}}]{Majumder2022}%
  \BibitemOpen
  \bibfield  {author} {\bibinfo {author} {\bibfnamefont {S.}~\bibnamefont
  {Majumder}}, \bibinfo {author} {\bibfnamefont {C.~G.}\ \bibnamefont {Yale}},
  \bibinfo {author} {\bibfnamefont {T.~D.}\ \bibnamefont {Morris}}, \bibinfo
  {author} {\bibfnamefont {D.~S.}\ \bibnamefont {Lobser}}, \bibinfo {author}
  {\bibfnamefont {A.~D.}\ \bibnamefont {Burch}}, \bibinfo {author}
  {\bibfnamefont {M.~N.}\ \bibnamefont {Chow}}, \bibinfo {author}
  {\bibfnamefont {M.~C.}\ \bibnamefont {Revelle}}, \bibinfo {author}
  {\bibfnamefont {S.~M.}\ \bibnamefont {Clark}}, \ and\ \bibinfo {author}
  {\bibfnamefont {R.~C.}\ \bibnamefont {Pooser}},\ }\href@noop {} {\bibfield
  {journal} {\bibinfo  {journal} {arXiv preprint arXiv:2205.14225}\ } (\bibinfo
  {year} {2022})}\BibitemShut {NoStop}%
\bibitem [{\citenamefont {Jones}(2003)}]{Jones2003}%
  \BibitemOpen
  \bibfield  {author} {\bibinfo {author} {\bibfnamefont {J.~A.}\ \bibnamefont
  {Jones}},\ }\href {\doibase 10.1103/PhysRevA.67.012317} {\bibfield  {journal}
  {\bibinfo  {journal} {Phys. Rev. A}\ }\textbf {\bibinfo {volume} {67}},\
  \bibinfo {pages} {012317} (\bibinfo {year} {2003})}\BibitemShut {NoStop}%
\bibitem [{\citenamefont {Merrill}\ and\ \citenamefont
  {Brown}(2014)}]{Merrill2014}%
  \BibitemOpen
  \bibfield  {author} {\bibinfo {author} {\bibfnamefont {J.~T.}\ \bibnamefont
  {Merrill}}\ and\ \bibinfo {author} {\bibfnamefont {K.~R.}\ \bibnamefont
  {Brown}},\ }\href@noop {} {\bibfield  {journal} {\bibinfo  {journal} {Adv.
  Chem. Phys}\ }\textbf {\bibinfo {volume} {154}},\ \bibinfo {pages} {241}
  (\bibinfo {year} {2014})}\BibitemShut {NoStop}%
\bibitem [{\citenamefont {Kang}\ \emph {et~al.}(2021)\citenamefont {Kang},
  \citenamefont {Liang}, \citenamefont {Zhang}, \citenamefont {Huang},
  \citenamefont {Wang}, \citenamefont {Fang}, \citenamefont {Kim},\ and\
  \citenamefont {Brown}}]{Kang2021}%
  \BibitemOpen
  \bibfield  {author} {\bibinfo {author} {\bibfnamefont {M.}~\bibnamefont
  {Kang}}, \bibinfo {author} {\bibfnamefont {Q.}~\bibnamefont {Liang}},
  \bibinfo {author} {\bibfnamefont {B.}~\bibnamefont {Zhang}}, \bibinfo
  {author} {\bibfnamefont {S.}~\bibnamefont {Huang}}, \bibinfo {author}
  {\bibfnamefont {Y.}~\bibnamefont {Wang}}, \bibinfo {author} {\bibfnamefont
  {C.}~\bibnamefont {Fang}}, \bibinfo {author} {\bibfnamefont {J.}~\bibnamefont
  {Kim}}, \ and\ \bibinfo {author} {\bibfnamefont {K.~R.}\ \bibnamefont
  {Brown}},\ }\href {\doibase 10.1103/PhysRevApplied.16.024039} {\bibfield
  {journal} {\bibinfo  {journal} {Phys. Rev. Applied}\ }\textbf {\bibinfo
  {volume} {16}},\ \bibinfo {pages} {024039} (\bibinfo {year}
  {2021})}\BibitemShut {NoStop}%
\bibitem [{\citenamefont {Webb}\ \emph {et~al.}(2018)\citenamefont {Webb},
  \citenamefont {Webster}, \citenamefont {Collingbourne}, \citenamefont
  {Bretaud}, \citenamefont {Lawrence}, \citenamefont {Weidt}, \citenamefont
  {Mintert},\ and\ \citenamefont {Hensinger}}]{Webb2018}%
  \BibitemOpen
  \bibfield  {author} {\bibinfo {author} {\bibfnamefont {A.~E.}\ \bibnamefont
  {Webb}}, \bibinfo {author} {\bibfnamefont {S.~C.}\ \bibnamefont {Webster}},
  \bibinfo {author} {\bibfnamefont {S.}~\bibnamefont {Collingbourne}}, \bibinfo
  {author} {\bibfnamefont {D.}~\bibnamefont {Bretaud}}, \bibinfo {author}
  {\bibfnamefont {A.~M.}\ \bibnamefont {Lawrence}}, \bibinfo {author}
  {\bibfnamefont {S.}~\bibnamefont {Weidt}}, \bibinfo {author} {\bibfnamefont
  {F.}~\bibnamefont {Mintert}}, \ and\ \bibinfo {author} {\bibfnamefont
  {W.~K.}\ \bibnamefont {Hensinger}},\ }\href {\doibase
  10.1103/PhysRevLett.121.180501} {\bibfield  {journal} {\bibinfo  {journal}
  {Phys. Rev. Lett.}\ }\textbf {\bibinfo {volume} {121}},\ \bibinfo {pages}
  {180501} (\bibinfo {year} {2018})}\BibitemShut {NoStop}%
\bibitem [{\citenamefont {Spivey}\ \emph {et~al.}(2022)\citenamefont {Spivey},
  \citenamefont {Inlek}, \citenamefont {Jia}, \citenamefont {Crain},
  \citenamefont {Sun}, \citenamefont {Kim}, \citenamefont {Vrijsen},
  \citenamefont {Fang}, \citenamefont {Fitzgerald}, \citenamefont {Kross},
  \citenamefont {Noel},\ and\ \citenamefont {Kim}}]{Spivey2022}%
  \BibitemOpen
  \bibfield  {author} {\bibinfo {author} {\bibfnamefont {R.~F.}\ \bibnamefont
  {Spivey}}, \bibinfo {author} {\bibfnamefont {I.~V.}\ \bibnamefont {Inlek}},
  \bibinfo {author} {\bibfnamefont {Z.}~\bibnamefont {Jia}}, \bibinfo {author}
  {\bibfnamefont {S.}~\bibnamefont {Crain}}, \bibinfo {author} {\bibfnamefont
  {K.}~\bibnamefont {Sun}}, \bibinfo {author} {\bibfnamefont {J.}~\bibnamefont
  {Kim}}, \bibinfo {author} {\bibfnamefont {G.}~\bibnamefont {Vrijsen}},
  \bibinfo {author} {\bibfnamefont {C.}~\bibnamefont {Fang}}, \bibinfo {author}
  {\bibfnamefont {C.}~\bibnamefont {Fitzgerald}}, \bibinfo {author}
  {\bibfnamefont {S.}~\bibnamefont {Kross}}, \bibinfo {author} {\bibfnamefont
  {T.}~\bibnamefont {Noel}}, \ and\ \bibinfo {author} {\bibfnamefont
  {J.}~\bibnamefont {Kim}},\ }\href {\doibase 10.1109/TQE.2021.3125926}
  {\bibfield  {journal} {\bibinfo  {journal} {IEEE Transactions on Quantum
  Engineering}\ }\textbf {\bibinfo {volume} {3}},\ \bibinfo {pages} {1}
  (\bibinfo {year} {2022})}\BibitemShut {NoStop}%
\bibitem [{\citenamefont {Ball}\ \emph {et~al.}(2021)\citenamefont {Ball},
  \citenamefont {Biercuk}, \citenamefont {Carvalho}, \citenamefont {Chen},
  \citenamefont {Hush}, \citenamefont {Castro}, \citenamefont {Li},
  \citenamefont {Liebermann}, \citenamefont {Slatyer}, \citenamefont {Edmunds},
  \citenamefont {Frey}, \citenamefont {Hempel},\ and\ \citenamefont
  {Milne}}]{Ball2021}%
  \BibitemOpen
  \bibfield  {author} {\bibinfo {author} {\bibfnamefont {H.}~\bibnamefont
  {Ball}}, \bibinfo {author} {\bibfnamefont {M.~J.}\ \bibnamefont {Biercuk}},
  \bibinfo {author} {\bibfnamefont {A.~R.~R.}\ \bibnamefont {Carvalho}},
  \bibinfo {author} {\bibfnamefont {J.}~\bibnamefont {Chen}}, \bibinfo {author}
  {\bibfnamefont {M.}~\bibnamefont {Hush}}, \bibinfo {author} {\bibfnamefont
  {L.~A.~D.}\ \bibnamefont {Castro}}, \bibinfo {author} {\bibfnamefont
  {L.}~\bibnamefont {Li}}, \bibinfo {author} {\bibfnamefont {P.~J.}\
  \bibnamefont {Liebermann}}, \bibinfo {author} {\bibfnamefont {H.~J.}\
  \bibnamefont {Slatyer}}, \bibinfo {author} {\bibfnamefont {C.}~\bibnamefont
  {Edmunds}}, \bibinfo {author} {\bibfnamefont {V.}~\bibnamefont {Frey}},
  \bibinfo {author} {\bibfnamefont {C.}~\bibnamefont {Hempel}}, \ and\ \bibinfo
  {author} {\bibfnamefont {A.}~\bibnamefont {Milne}},\ }\href {\doibase
  10.1088/2058-9565/abdca6} {\bibfield  {journal} {\bibinfo  {journal} {Quantum
  Science and Technology}\ }\textbf {\bibinfo {volume} {6}},\ \bibinfo {pages}
  {044011} (\bibinfo {year} {2021})}\BibitemShut {NoStop}%
\bibitem [{\citenamefont {Revelle}(2020)}]{Revelle2020}%
  \BibitemOpen
  \bibfield  {author} {\bibinfo {author} {\bibfnamefont {M.~C.}\ \bibnamefont
  {Revelle}},\ }\href@noop {} {\bibfield  {journal} {\bibinfo  {journal} {arXiv
  preprint arXiv:2009.02398}\ } (\bibinfo {year} {2020})}\BibitemShut {NoStop}%
\bibitem [{\citenamefont {Lobser}\ \emph {et~al.}(2020)\citenamefont {Lobser},
  \citenamefont {Maunz}, \citenamefont {Van Der~Wall},\ and\ \citenamefont
  {USDOE}}]{RFSoC}%
  \BibitemOpen
  \bibfield  {author} {\bibinfo {author} {\bibfnamefont {D.}~\bibnamefont
  {Lobser}}, \bibinfo {author} {\bibfnamefont {P.}~\bibnamefont {Maunz}},
  \bibinfo {author} {\bibfnamefont {J.}~\bibnamefont {Van Der~Wall}}, \ and\
  \bibinfo {author} {\bibnamefont {USDOE}},\ }\href {\doibase
  10.11578/dc.20201112.1} {\enquote {\bibinfo {title} {Octet: A coherent qubit
  control system for trapped-ion quantum computers},}\ } (\bibinfo {year}
  {2020})\BibitemShut {NoStop}%
\bibitem [{\citenamefont {Leibfried}\ \emph {et~al.}(2003)\citenamefont
  {Leibfried}, \citenamefont {DeMarco}, \citenamefont {Meyer}, \citenamefont
  {Lucas}, \citenamefont {Barrett}, \citenamefont {Britton}, \citenamefont
  {Itano}, \citenamefont {Jelenkovi{\'c}}, \citenamefont {Langer},
  \citenamefont {Rosenband} \emph {et~al.}}]{Leibfried2003}%
  \BibitemOpen
  \bibfield  {author} {\bibinfo {author} {\bibfnamefont {D.}~\bibnamefont
  {Leibfried}}, \bibinfo {author} {\bibfnamefont {B.}~\bibnamefont {DeMarco}},
  \bibinfo {author} {\bibfnamefont {V.}~\bibnamefont {Meyer}}, \bibinfo
  {author} {\bibfnamefont {D.}~\bibnamefont {Lucas}}, \bibinfo {author}
  {\bibfnamefont {M.}~\bibnamefont {Barrett}}, \bibinfo {author} {\bibfnamefont
  {J.}~\bibnamefont {Britton}}, \bibinfo {author} {\bibfnamefont {W.~M.}\
  \bibnamefont {Itano}}, \bibinfo {author} {\bibfnamefont {B.}~\bibnamefont
  {Jelenkovi{\'c}}}, \bibinfo {author} {\bibfnamefont {C.}~\bibnamefont
  {Langer}}, \bibinfo {author} {\bibfnamefont {T.}~\bibnamefont {Rosenband}},
  \emph {et~al.},\ }\href@noop {} {\bibfield  {journal} {\bibinfo  {journal}
  {Nature}\ }\textbf {\bibinfo {volume} {422}},\ \bibinfo {pages} {412}
  (\bibinfo {year} {2003})}\BibitemShut {NoStop}%
\bibitem [{\citenamefont {Solano}\ \emph {et~al.}(1999)\citenamefont {Solano},
  \citenamefont {de~Matos~Filho},\ and\ \citenamefont {Zagury}}]{Solano1999}%
  \BibitemOpen
  \bibfield  {author} {\bibinfo {author} {\bibfnamefont {E.}~\bibnamefont
  {Solano}}, \bibinfo {author} {\bibfnamefont {R.~L.}\ \bibnamefont
  {de~Matos~Filho}}, \ and\ \bibinfo {author} {\bibfnamefont {N.}~\bibnamefont
  {Zagury}},\ }\href {\doibase 10.1103/PhysRevA.59.R2539} {\bibfield  {journal}
  {\bibinfo  {journal} {Phys. Rev. A}\ }\textbf {\bibinfo {volume} {59}},\
  \bibinfo {pages} {R2539} (\bibinfo {year} {1999})}\BibitemShut {NoStop}%
\bibitem [{\citenamefont {Milburn}\ \emph {et~al.}(2000)\citenamefont
  {Milburn}, \citenamefont {Schneider},\ and\ \citenamefont
  {James}}]{Milburn2000}%
  \BibitemOpen
  \bibfield  {author} {\bibinfo {author} {\bibfnamefont {G.}~\bibnamefont
  {Milburn}}, \bibinfo {author} {\bibfnamefont {S.}~\bibnamefont {Schneider}},
  \ and\ \bibinfo {author} {\bibfnamefont {D.}~\bibnamefont {James}},\
  }\href@noop {} {\bibfield  {journal} {\bibinfo  {journal} {Fortschritte der
  Physik: Progress of Physics}\ }\textbf {\bibinfo {volume} {48}},\ \bibinfo
  {pages} {801} (\bibinfo {year} {2000})}\BibitemShut {NoStop}%
\bibitem [{\citenamefont {Shapira}\ \emph {et~al.}(2022)\citenamefont
  {Shapira}, \citenamefont {Cohen}, \citenamefont {Akerman}, \citenamefont
  {Stern},\ and\ \citenamefont {Ozeri}}]{Shapira2022}%
  \BibitemOpen
  \bibfield  {author} {\bibinfo {author} {\bibfnamefont {Y.}~\bibnamefont
  {Shapira}}, \bibinfo {author} {\bibfnamefont {S.}~\bibnamefont {Cohen}},
  \bibinfo {author} {\bibfnamefont {N.}~\bibnamefont {Akerman}}, \bibinfo
  {author} {\bibfnamefont {A.}~\bibnamefont {Stern}}, \ and\ \bibinfo {author}
  {\bibfnamefont {R.}~\bibnamefont {Ozeri}},\ }\href@noop {} {\bibfield
  {journal} {\bibinfo  {journal} {arXiv preprint arXiv:2207.01660}\ } (\bibinfo
  {year} {2022})}\BibitemShut {NoStop}%
\bibitem [{\citenamefont {Ruzic}\ \emph {et~al.}(2022)\citenamefont {Ruzic},
  \citenamefont {Chow}, \citenamefont {Burch}, \citenamefont {Lobser},
  \citenamefont {Revelle}, \citenamefont {Wilson}, \citenamefont {Yale},\ and\
  \citenamefont {Clark}}]{Ruzic2022}%
  \BibitemOpen
  \bibfield  {author} {\bibinfo {author} {\bibfnamefont {B.~P.}\ \bibnamefont
  {Ruzic}}, \bibinfo {author} {\bibfnamefont {M.~N.~H.}\ \bibnamefont {Chow}},
  \bibinfo {author} {\bibfnamefont {A.~D.}\ \bibnamefont {Burch}}, \bibinfo
  {author} {\bibfnamefont {D.}~\bibnamefont {Lobser}}, \bibinfo {author}
  {\bibfnamefont {M.~C.}\ \bibnamefont {Revelle}}, \bibinfo {author}
  {\bibfnamefont {J.~M.}\ \bibnamefont {Wilson}}, \bibinfo {author}
  {\bibfnamefont {C.~G.}\ \bibnamefont {Yale}}, \ and\ \bibinfo {author}
  {\bibfnamefont {S.~M.}\ \bibnamefont {Clark}},\ }\href@noop {} {\bibfield
  {journal} {\bibinfo  {journal} {arXiv preprint arXiv:2210.02372}\ } (\bibinfo
  {year} {2022})}\BibitemShut {NoStop}%
\end{thebibliography}%

\end{document}